\documentclass[journal]{IEEEtran}
\usepackage[noadjust]{cite}
\usepackage{srcltx}
\usepackage{graphicx}
\usepackage{epstopdf}
\usepackage{psfrag}
\usepackage{epsfig}
\usepackage[tbtags,sumlimits,nointlimits,reqno]{amsmath}
\usepackage{amssymb}
\usepackage{xcolor}
\usepackage{float}
\usepackage{subfigure}
\usepackage[most]{tcolorbox}
\usepackage{soul}
\usepackage{footnote}
\usepackage{amsmath}
\usepackage{color}

\usepackage[framemethod=TikZ]{mdframed}
\usepackage{lipsum}
\mdfdefinestyle{MyFrame}{%
	linecolor=blue,
	outerlinewidth=2pt,
	roundcorner=20pt,
	innertopmargin=\baselineskip,
	innerbottommargin=\baselineskip,
	innerrightmargin=20pt,
	innerleftmargin=20pt,
	backgroundcolor=gray!50!white}

\begin{document}
	\title{Light Interaction With a Space-Time-Modulated Josephson Junction Array and Application to Angular-Frequency Beam Multiplexing}
	\author{Sajjad Taravati,~\IEEEmembership{Senior Member,~IEEE}
		\thanks{S. Taravati is with the Faculty of Engineering and Physical Sciences, University of Southampton, Southampton SO17 1BJ, UK (e-mail: s.taravati@soton.ac.uk).}%
		\thanks{Manuscript received ***, 2025; revised ***, 2025.}}
	
	\markboth{}%
	{*** \MakeLowercase{\textit{et al.}}: Bare Demo of IEEEtran.cls for IEEE Journals}

	\maketitle
	
	\begin{abstract}
Josephson junctions, as pivotal components of modern technologies such as superconducting quantum computing, owe their prominence to their unique nonlinear properties at low temperatures. Despite their extensive use in static configurations, the study of dynamic Josephson junctions, particularly under space-time modulation, remains largely unexplored. This study investigates the interaction and transmission of electromagnetic waves through arrays of space-time-modulated Josephson junctions. A comprehensive mathematical framework is presented to model the propagation of electric and magnetic fields within and beyond these structures. We demonstrate how such dynamic arrays enable groundbreaking four-dimensional light manipulation, achieving angular-frequency beam multiplexing through a seamless integration of frequency conversion and beam-splitting functionalities. These advancements open new horizons for electromagnetic field engineering, with far-reaching implications for superconducting quantum technologies, next-generation wireless communications, biomedical sensing, and radar systems.
	\end{abstract}
	
	
	\IEEEpeerreviewmaketitle

\section{Introduction}\label{sec:introduction}
\IEEEPARstart{J}osephson junctions, formed by sandwiching an insulating layer between two superconducting layers, are pivotal components in superconducting electronics~\cite{makhlin2001quantum,kleiner2021space}. These \textit{nonlinear} devices exploit the quantum mechanical phenomenon of supercurrent—a current that flows without resistance across the insulator via Cooper pair tunneling. Their unique ability to manipulate electromagnetic waves at the quantum level has made them indispensable in a broad spectrum of applications, ranging from classical electronics to quantum technologies. In superconducting quantum technologies, Josephson junctions serve as the foundational building blocks for qubits in quantum computing, enabling high-coherence quantum state manipulation. Their nonlinearity and tunability make them ideal for designing parametric amplifiers, quantum-limited detectors, and frequency converters in ultra-sensitive measurement systems. Furthermore, Josephson junctions have revolutionized precision metrology, providing the basis for voltage standards and microwave generation with unparalleled accuracy. Beyond quantum computing and metrology, these devices are being actively explored in the development of quantum-limited sensors for astrophysics, advanced communication systems leveraging quantum states, and scalable superconducting circuits for quantum information processing.

Despite their established applications, the integration of Josephson junctions into space-time-modulated systems remains largely unexplored~\cite{taravati2024spatiotemporal,taravati2024efficient,taravati2024one}. Our research represents one of the first efforts to investigate the interaction of electromagnetic waves with space-time-modulated Josephson junction arrays, marking a significant advancement in this emerging field. By dynamically modulating the properties of Josephson junctions in both space and time, we open up new opportunities for achieving nonreciprocal light propagation, tunable frequency conversion, and tailored photon interactions. These capabilities hold transformative potential for quantum communication, photon-based information processing, and other cutting-edge quantum technologies.

Over the past decade, linear space-time-modulated structures have attracted significant attention due to their wide-ranging applications in modern wireless communication systems, photonics, and radar technologies~\cite{Fan_NPH_2009,Taravati_Kishk_PRB_2018,Taravati_PRAp_2018,Taravati_Kishk_MicMag_2019,wang2022imaging,saikia2022time,Taravati_Kishk_TAP_2019,taravati2020full,solis2021functional,pacheco2021temporal,taravati2024finite}. Such arrays exhibit dynamic properties characterized by the modulation of electrical permittivity across both space and time~\cite{wang2021space,wan2021nonreciprocal,taravati20234d,amra2024linear,Taravati_NC_2021,valizadeh2024analytical}, yielding linear spatiotemporal metasurfaces at microwave~\cite{saikia2019frequency,Taravati_ACSP_2022,kumar2024multi} and optical frequencies~\cite{salary2020time,sabri2021broadband,sisler2024electrically}. These efforts have led to introduction of original high efficiency functionalities and devices such as nonreciprocal transmission~\cite{taravati2020full,cardin2020surface}, target recognition~\cite{wang2023pseudorandom}, isolators~\cite{lira2012electrically,Taravati_PRB_SB_2017,Taravati_AMTech_2021}, temporal aiming~\cite{pacheco2020temporal}, frequency conversion~\cite{taravati2021pure,Taravati_PRB_Mixer_2018}, static-to-dynamic field conversion~\cite{mencagli2022static}, circulators~\cite{dinc2017millimeter,Alu_NPH_2014}, parametric amplification~\cite{zhu2020tunable,pendry2020new}, multiple access secure communication systems~\cite{taravati_PRApp_2019,sedeh2021active}, nonreciprocal antennas~\cite{zang2019nonreciprocal}, and multifunctional operations~\cite{Taravati_AMA_PRApp_2020,wang2019multifunctional}. Conventional time-modulated devices and frequency converters, along with their electronic components such as varactors, transistors, and diodes, are poorly suited for the demanding millikelvin-temperature environments of superconducting quantum technologies. These components suffer from significant operational constraints and introduce undesirable noise, rendering them incompatible with the stringent performance requirements of quantum systems. In space-time-modulated media, the governing wave equation gives rise to coupled differential equations for each harmonic. At low modulation frequencies, strong coupling between harmonics occurs, enabling efficient sideband generation. In nonlinear systems, higher-order terms become dominant, facilitating effective harmonic generation even at high modulation frequencies. By harnessing these properties, space-time-modulated Josephson junction arrays achieve highly efficient frequency conversion, even at large modulation-to-signal frequency ratios. Unlike linear STM systems, which are constrained by factors such as phase matching, dispersion, and the need for long interaction lengths, nonlinear STM Josephson arrays deliver superior performance in a compact form factor. This compactness, combined with high efficiency and precision, makes them particularly well-suited for superconducting quantum technologies operating at millikelvin temperatures, where these attributes are critical for optimal functionality.

Furthermore, this study shows that an array of dynamic Josephson junctions presents angular-frequency beam multiplexing. Frequency conversion and beam-splitting are two pivotal operations for advancing cutting-edge technologies such as quantum technologies and wireless communications. These functionalities enable the efficient transfer and manipulation of quantum information and communication signals across different frequency bands~\cite{albrecht2014waveguide,bock2018high,maring2018quantum,han2021microwave,santiago2022resonant,ataloglou2023metasurfaces,geus2024low}. In quantum computing, for instance, frequency converters are essential for mediating interactions between qubits operating at different frequencies to reduce crosstalk and enhance qubit fidelity—critical for implementing robust quantum error correction protocols~\cite{gambetta2017building,he2022control}. Furthermore, they facilitate the transduction of quantum states between microwave photons and optical photons, which are optimal for long-distance quantum communication and integration with long-lived quantum memories. This capability is crucial for extending quantum computing networks and creating hybrid quantum systems~\cite{han2021microwave}. In quantum sensing, frequency converters operating at millikelvin temperatures with ultra-low noise significantly enhance sensitivity and precision, which enables advanced applications such as dark matter detection and gravitational wave sensing~\cite{berlin2020axion,dixit2021searching,bass2024quantum}. Similarly, in wireless communications, frequency conversion is integral to modern systems. It enables the bridging of different frequency bands, which is vital for seamless connectivity in multiband networks, such as 5G and beyond. For example, frequency converters are used to integrate millimeter-wave bands with lower-frequency sub-6 GHz bands, optimizing bandwidth utilization and ensuring backward compatibility in heterogeneous networks. Beam-splitting, on the other hand, facilitates multi-user access and signal distribution in massive multiple-input multiple-output (MIMO) systems~\cite{dubuc2004mimo}. This leads to an enhanced spectral efficiency and communication reliability. Additionally, both operations play a key role in satellite and airborne communication systems, where efficient frequency management and beam shaping are essential for maintaining stable and high-speed links across diverse environments.

This paper is organized as follows: Section~\ref{sec:theory} explores the theoretical framework and mathematical modeling of light interaction with space-time-modulated Josephson junction arrays. Section~\ref{sec:inc} examines the incidence and transmission of light through these dynamic arrays, detailing the associated mechanisms. Section~\ref{sec:exp} proposes a practical implementation of the space-time-modulated Josephson junction array and presents full-wave numerical simulation results that demonstrate its frequency-angular beam multiplexing capabilities. Finally, Section~\ref{sec:conc} summarizes the findings and concludes the paper.

\section{Theoretical Implications}\label{sec:theory}
\subsection{Array of space-time-modulated Josephson junctions}
\begin{figure*}
	\begin{center}
		\includegraphics[width=1.5\columnwidth]{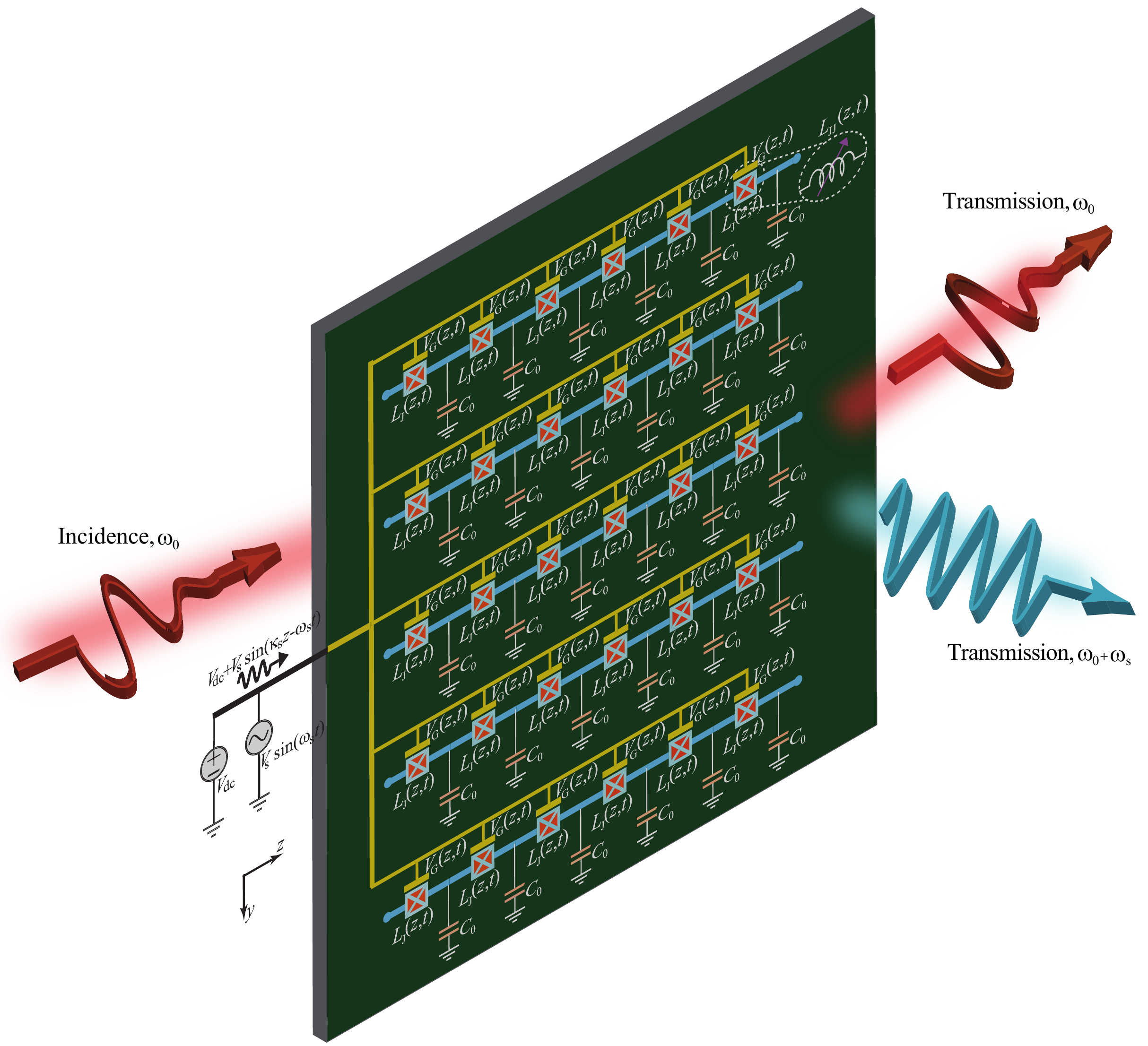}
		\caption{Light interaction with an array of space-time modulated Josephson junctions leading to angular-frequency beam multiplexing.}
		\label{Fig:sch}
	\end{center}
\end{figure*}

The structure under study, shown in Fig.~\ref{Fig:sch}, consists of an array of space-time-modulated Josephson junctions, with the current density of the array expressed as
\begin{subequations}
\begin{equation}
J(z,t)=I_0 \sin[\rho(z,t)],
\end{equation}
where $I_0$ is the maximum current and
\begin{equation}\label{eqa:rho}
\rho(z,t)=2\pi \dfrac{\Phi(z,t)}{\Phi_0},
\end{equation}
where $\Phi_0$ is the magnetic flux quantum and $\Phi(z,t)$ is the time- and space-dependent magnetic flux given by
\begin{equation}
\Phi(z,t)=\int_{-\infty}^t V(z,t') dt'.
\end{equation}

The time derivative of the current density, which is directly related to the voltage $V(z,t)$, is given by
\begin{equation}
	\dfrac{d J(z,t)}{dt}=\dfrac{2\pi I_0 \cos[\rho(z,t)]}{\Phi_0} V(z,t)
\end{equation}
which gives
\begin{equation}
	V(z,t)=L(z,t) \dfrac{d J(z,t)}{dt}
\end{equation}
where
\begin{equation}\label{eqa:L1}
L(z,t)=\dfrac{\Phi_0}{2\pi I_0 \cos[\rho(z,t)]},
\end{equation}

Therefore, $\rho(z,t)$ given by Eq.~\eqref{eqa:rho} reads
\begin{equation}
		\begin{split}
\rho(z,t)&=\dfrac{2\pi}{\Phi_0} \int_{-\infty}^t V(z,t') dt'\\
&=\dfrac{2\pi}{\Phi_0} \left( t V_\text{dc}+\dfrac{1}{\omega} V_\text{rf} \sin[\kappa_\text{s} z-\omega_\text{s} t+\phi] \right),
	\end{split}	
\end{equation}

As a result, the space-time-varying inductance given by Eq.~\eqref{eqa:L1} is expressed as
\begin{equation}\label{eqa:L2}
	L_\text{S}(z,t)= \dfrac{\Phi_0}{2\pi I_0 \cos\left(\widetilde{\Phi}_\text{dc}+ \widetilde{\Phi}_\text{rf} \sin[\kappa_\text{s} z-\omega_\text{s} t+\phi]\right) },
\end{equation}
\end{subequations}
where $\widetilde{\Phi}_\text{dc}= 2\pi \Phi_\text{dc}/ \Phi_0$ and $\widetilde{\Phi}_\text{rf}= 2\pi \Phi_\text{rf}/ \Phi_0$. Consequently, the effective space-time-varying magnetic permeability of the array is expressed as
\begin{equation}\label{eqa:ap_permeab}
\begin{split}
	&\mu_\text{s}(z,t)= \dfrac{l L_\text{S}(z,t)}{\mu_0A} 
	= \dfrac{1}{  G_{\mu} \cos\left(\widetilde{\Phi}_\text{dc}+ \widetilde{\Phi}_\text{rf} \sin[\kappa_\text{s} z-\omega_\text{s} t+\phi]\right)},
\end{split}
\end{equation}

where $G_{\mu}=2\pi \mu_0 I_0 A/(\Phi_0 l)$, where $l$ and $A$ are the length and area of the space-time-modulated array, respectively.

Since the array is periodic in both space and time, the magnetic permeability of the array may be decomposed into space-time Floquet-Bloch harmonics, as
\begin{subequations}
\begin{equation}\label{eqa:g}
	g(z,t)=  \dfrac{1}{\mu_\text{s}(z,t)}=\sum_{m=-\infty}^{+\infty} g_m e^{-jm(\kappa_\text{s} z-\omega_\text{s} t+\phi)} ,
\end{equation}

To find the unknown coefficients $g_m$s we first define $\psi=\kappa_\text{s} z-\omega_\text{s} t+\phi$, and then expand the cosine term in Eq.~\eqref{eqa:ap_permeab}, as
\begin{equation}\label{eqa:Exp}
\begin{split}
\cos\left(\widetilde{\Phi}_\text{dc}+\widetilde{\Phi}_\text{rf} \sin \psi \right)
 &=\cos\left( \widetilde{\Phi}_\text{dc} \right) \cos\left( \widetilde{\Phi}_\text{rf} \sin \psi\right)
\\& -\sin\left( \widetilde{\Phi}_\text{dc} \right) \sin\left( \widetilde{\Phi}_\text{rf} \sin \psi\right).
\end{split}	
\end{equation}

The cosine and sine terms in Eq.~\eqref{eqa:Exp} can be further expanded using the Taylor's series expansion, as
\begin{equation}\label{eqa:T1}
	\begin{split}
\cos&\left( \widetilde{\Phi}_\text{rf} \sin \psi\right)=
	\left(1-\dfrac{{\widetilde{\Phi}_\text{rf}^2}}{4}+\dfrac{{\widetilde{\Phi}_\text{rf}^4}}{64}-\dfrac{{\widetilde{\Phi}_\text{rf}^6}}{2304} +\dfrac{{\widetilde{\Phi}_\text{rf}^8}}{147456}\right)\\&+\cos(2\psi) \left( \dfrac{{\widetilde{\Phi}_\text{rf}^2}}{4}- \dfrac{{\widetilde{\Phi}_\text{rf}^4}}{48}  +\dfrac{{\widetilde{\Phi}_\text{rf}^6}}{1536} - \dfrac{{\widetilde{\Phi}_\text{rf}^8}}{92160}\right) \\&+\cos(4\psi)
	\left(  \dfrac{{\widetilde{\Phi}_\text{rf}^4}}{192}  - \dfrac{{\widetilde{\Phi}_\text{rf}^6}}{3840} + \dfrac{{\widetilde{\Phi}_\text{rf}^8}}{184320}\right) \\&
	+\cos(6\psi)
	\left(  \dfrac{{\widetilde{\Phi}_\text{rf}^6}}{23040} - \dfrac{{\widetilde{\Phi}_\text{rf}^8}}{645120}\right)
		+\cos(8\psi)
  \dfrac{{\widetilde{\Phi}_\text{rf}^8}}{5160960},
\end{split}
\end{equation}
and
\begin{equation}\label{eqa:T2}
\begin{split}
\sin&\left( \widetilde{\Phi}_\text{rf} \sin \psi\right)=\sin\psi \left(\widetilde{\Phi}_\text{rf}-\dfrac{{\widetilde{\Phi}_\text{rf}^3}}{8}+\dfrac{{\widetilde{\Phi}_\text{rf}^5}}{192}-\dfrac{{\widetilde{\Phi}_\text{rf}^7}}{9216}  \right)   \\&
+\sin(3\psi) \left( \dfrac{{\widetilde{\Phi}_\text{rf}^3}}{24}-\dfrac{{\widetilde{\Phi}_\text{rf}^5}}{384}+\dfrac{{\widetilde{\Phi}_\text{rf}^7}}{15360}  \right) \\&
+\sin(5\psi) \left( \dfrac{{\widetilde{\Phi}_\text{rf}^5}}{1920}-\dfrac{{\widetilde{\Phi}_\text{rf}^7}}{46080}  \right)
-\sin(7\psi)  \dfrac{{\widetilde{\Phi}_\text{rf}^7}}{322560}.
\end{split}
\end{equation}

Therefore, the $g_m$s in Eq.~\eqref{eqa:g} read
\begin{equation}
\begin{split}
&	g_0=G_{\mu} \cos\left( \widetilde{\Phi}_\text{dc} \right) 	\left(1-\dfrac{{\widetilde{\Phi}_\text{rf}^2}}{4}+\dfrac{{\widetilde{\Phi}_\text{rf}^4}}{64}-\dfrac{{\widetilde{\Phi}_\text{rf}^6}}{2304} +\dfrac{{\widetilde{\Phi}_\text{rf}^8}}{147456}\right)\\&
g_1=-g_{-1}=\dfrac{G_{\mu} }{2j}\sin\left( \widetilde{\Phi}_\text{dc} \right) \left(\widetilde{\Phi}_\text{rf}-\dfrac{{\widetilde{\Phi}_\text{rf}^3}}{8}+\dfrac{{\widetilde{\Phi}_\text{rf}^5}}{192}-\dfrac{{\widetilde{\Phi}_\text{rf}^7}}{9216}  \right)  \\&
	g_2=g_{-2}= \dfrac{G_{\mu}}{8}  \cos\left( \widetilde{\Phi}_\text{dc} \right) \left( {\widetilde{\Phi}_\text{rf}^2}- \dfrac{{\widetilde{\Phi}_\text{rf}^4}}{12}  +\dfrac{{\widetilde{\Phi}_\text{rf}^6}}{384} - \dfrac{{\widetilde{\Phi}_\text{rf}^8}}{23040}\right	)\\&
	g_3=-g_{-3}=\dfrac{G_{\mu} }{j48}\sin\left( \widetilde{\Phi}_\text{dc} \right) 
	\left( {\widetilde{\Phi}_\text{rf}^3}-\dfrac{{\widetilde{\Phi}_\text{rf}^5}}{16}+\dfrac{{\widetilde{\Phi}_\text{rf}^7}}{640}  \right)
  \\&
  	g_4=g_{-4}= \dfrac{G_{\mu}}{384}  \cos\left( \widetilde{\Phi}_\text{dc} \right)
  	\left(  {\widetilde{\Phi}_\text{rf}^4}  - \dfrac{{\widetilde{\Phi}_\text{rf}^6}}{20} + \dfrac{{\widetilde{\Phi}_\text{rf}^8}}{960}\right)   \\&
  		g_5=-g_{-5}=\dfrac{G_{\mu} }{j3840}\sin\left( \widetilde{\Phi}_\text{dc} \right) \left( {\widetilde{\Phi}_\text{rf}^5}-\dfrac{{\widetilde{\Phi}_\text{rf}^7}}{24}  \right)  \\&
  g_6=g_{-6}= \dfrac{G_{\mu}}{46080}  \cos\left( \widetilde{\Phi}_\text{dc} \right)
  			\left(  {\widetilde{\Phi}_\text{rf}^6} - \dfrac{{\widetilde{\Phi}_\text{rf}^8}}{28}\right)  \\&
  		g_7=-g_{-7}=\dfrac{G_{\mu} }{j645120}\sin\left( \widetilde{\Phi}_\text{dc} \right) {\widetilde{\Phi}_\text{rf}^7}  \\&
 g_8=g_{-8}= \dfrac{G_{\mu}}{10321920}  \cos\left( \widetilde{\Phi}_\text{dc} \right)
  		  {\widetilde{\Phi}_\text{rf}^8}		
\end{split}
\end{equation}
\end{subequations}

\subsection{Floquet-Bloch Space-Time Harmonics}

\begin{figure}
	\begin{center}
		\includegraphics[width=1\columnwidth]{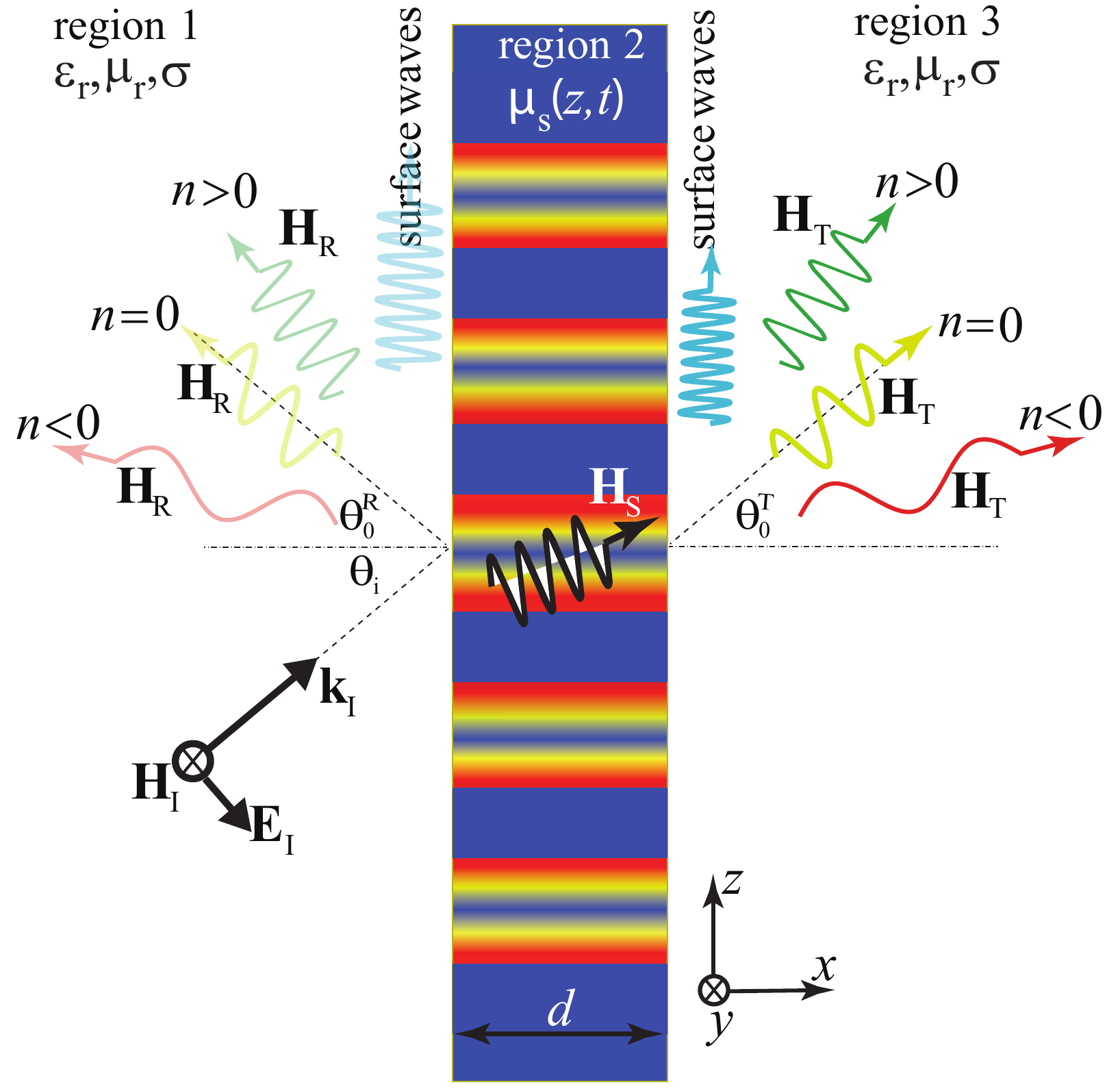}
		\caption{Wave scattering from a space-time superconducting surface.}
		\label{Fig:slab}
	\end{center}
\end{figure}

Figure~\ref{Fig:slab} illustrates the transmission and reflection of light from an array of space-time-periodic Josephson junctions, characterized by the space-time-periodic permeability described in Eq.~\eqref{eqa:ap_permeab}. aking into account the spatiotemporal periodicity of the spatiotemporal array, the electric and magnetic fields within the slab can be decomposed into a series of Floquet-Bloch harmonics, capturing the contributions of both spatial and temporal modulations. This expansion provides a comprehensive framework to describe wave interactions in periodic structures. Consequently, the transverse magnetic (TM) wave propagation within the slab can be expressed in terms of these harmonics, where each harmonic corresponds to a distinct combination of spatial and temporal frequencies. This representation not only facilitates the analysis of wave dispersion and mode coupling but also enables the study of energy redistribution among harmonics due to the spatiotemporal modulation. Such insights are particularly critical for understanding nonreciprocal wave dynamics, frequency conversion, and the emergence of higher-order harmonic modes, which are intrinsic to space-time-modulated systems. By expanding the fields in this manner, the underlying physics of wave propagation, including interactions between the primary and secondary harmonics, can be accurately captured and analyzed. Consequently, the transverse magnetic (TM) wave propagation within the slab can be expressed as
\begin{subequations}
	\begin{equation}
		\begin{split}
			\mathbf{H}_\text{s}(x,z,t)&=\mathbf{\hat{y}}\sum_{n }   H_{n}  e^{ - j \left[ k_x x+ \kappa_{n} z -\omega_n t\right]} ,
		\end{split}
		\label{eqa:A-E_mod_field_TM}
	\end{equation}
	and
	\begin{equation}
		\begin{split}
			&\mathbf{E}_\text{s}(x,z,t)=
			-\eta_2\left[\mathbf{\hat{k}}_\text{s}  \times \mathbf{H}_\text{s} (x,z,t)\right]\\
			&=\eta_2 \sum_{n }   \left(\mathbf{\hat{x}} \sin(\theta_{n}) - \mathbf{\hat{z}} \cos(\theta_\text{i}) \right)   H_{n}  e^{ - j \big[k_x x+ \kappa_{n}z -\omega_n t\big]} .
		\end{split}
		\label{eqa:A-H_mod_field}
	\end{equation}
\end{subequations}
where $k_x=k_0\cos(\theta_\text{i})$, $\theta_\text{i}$ is the angle between the incident wave and the array boundary, $\sin(\theta_n) = \kappa_n / k_n$, $\mathbf{\hat{k}}_\text{s} = \mathbf{\hat{x}} \cos(\theta_\text{i}) + \mathbf{\hat{z}} \kappa_n / k_n$, and $\kappa_n = \kappa_0 + n \kappa_\text{s}$ is the $z$-component of the wavenumber for the $n$th space-time harmonic inside the array. The source-free wave equation for the system is
\begin{subequations}
\begin{equation}
	\nabla^2 \mathbf{H}_\text{s}(x,z,t) - \frac{1}{{{c^2}}}\frac{{{\partial ^2} \left[\mu_\text{s} (z,t)\mathbf{H}_\text{s}(x,z,t) \right]}}{{\partial {t^2}}}=0.
	\label{eqa:wave_equ}
\end{equation}

Substituting the magnetic field expression from Eq.~\eqref{eqa:A-E_mod_field_TM} into the wave equation~\eqref{eqa:wave_equ} yields
\begin{equation}
	\begin{split}
		&\sum_{n}(k_x^2+ \kappa_n^2)	   H_{n}  e^{ - j \left[ k_x x+ \kappa_{n} z -\omega_n t\right]} \\&-\dfrac{1}{c^2} \dfrac{\partial ^2}{\partial {t^2}}  \sum_{m,n}  \widetilde{\mu}_m H_{m+n}  e^{ - j \left[ k_x x+ \kappa_{n} z -\omega_n t+\phi\right]} =0,
	\end{split}
	\label{eqa:mmmm}
\end{equation}
which may be cast as
\begin{equation}
	H_{n} e^{j\phi} \left[ \frac{ k_x^2+ \kappa_n^2  }{k_n^2 }\right] 
	-  \sum\limits_m   \tilde{\mu}_m H_{m+n}  =0,
	\label{eqa:recurs_gen}
\end{equation}
\end{subequations}
where $\tilde{\mu}_m=1/g_m$. Truncating to $2N+1$ terms yields  
\begin{subequations}
	\begin{equation}	\label{eqa:Equation}
		[\textbf{U}]\cdot[\overrightarrow{H}]=0.
	\end{equation}

The matrix $[\textbf{U}]$ is a square matrix of size $(2N+1) \times (2N+1)$, with elements defined as
	\begin{equation}
		\begin{split}
			U_{nn} &= \frac{ k_x^2+ \kappa_n ^2  }{k_n^2 }-  \tilde{\mu}_0,   \\
			U_{nm} &= -  \tilde{\mu}_{m+n},\quad\text{for }n\neq m.
		\end{split}
		\label{eqa:A_matrix}
	\end{equation}
where
	\begin{equation}\label{eqa:matr}
				\begin{split}
	&	[\textbf{U}]=\\	&\begin{bmatrix}
			c_{-N}  &  \tilde{\mu}_1 & \tilde{\mu}_2& \cdots  & \tilde{\mu}_{M-2}&\tilde{\mu}_{M-1}&\tilde{\mu}_M\\
			\tilde{\mu}_{-1}  &  c_{-N+1} & \tilde{\mu}_{1} &\cdots & \tilde{\mu}_{M-3}&\tilde{\mu}_{M-2}&\tilde{\mu}_{M-1}\\
			\tilde{\mu}_{-2}&\tilde{\mu}_{-1} &  c_{-N+2} & \cdots & \tilde{\mu}_{M-4}&\tilde{\mu}_{M-3}&\tilde{\mu}_{M-2}\\
			\vdots        &    \vdots    &    \vdots  &    \ddots  & \vdots  &        \vdots          &      \vdots\\
			\tilde{\mu}_{-M+2} 	&\tilde{\mu}_{-M+3} & \tilde{\mu}_{-M+4} &\cdots &	 c_{N-2} & \tilde{\mu}_{1}&\tilde{\mu}_{2}\\
			\tilde{\mu}_{-M+1} 	&\tilde{\mu}_{-M+2} & \tilde{\mu}_{-M+3} &\cdots &	\tilde{\mu}_{-1}  &  c_{N-1} & \tilde{\mu}_{1}\\
			\tilde{\mu}_{-M}  &  \tilde{\mu}_{-M+1} & \tilde{\mu}_{-M+2} &\cdots   &	\tilde{\mu}_{-2}&\tilde{\mu}_{-1} & c_{N}\\
		\end{bmatrix},
	\end{split}
	\end{equation}
	where $c_n=\tilde{\mu}_0-(k_x^2+ \kappa_n ^2)/k_n^2$. The vector of unknowns $[\overrightarrow{H}]$ in Eq.~\eqref{eqa:Equation} is a $(2N+1)\times 1$ vector containing the $H_n$ values, given by
	\begin{equation}\label{eqa:matr2}
		\overrightarrow{H}=	
		\begin{bmatrix}
			H_{-N}\\
			H_{-N+1}\\
			H_{-N+2} \\
			\vdots     \\
			H_{N-2}\\
			H_{N-1} \\
			H_{N}\\
		\end{bmatrix},
	\end{equation}

For non-trivial solutions (i.e., $[\overrightarrow{H}] \neq 0$), the matrix $[\textbf{U}]$ must be singular, meaning its determinant is zero, that is,
\begin{equation}
	\det\left[\textbf{U}\right]=0,
	\label{eqa:det_gen}
\end{equation}
\end{subequations}
\indent The condition outlined in Eq.~\eqref{eqa:det_gen} determines when the system permits wave propagation, offering the dispersion relationship typically expressed as \(\omega(k)\) or \(k(\omega)\). The matrix \([\textbf{U}]\) encapsulates material properties, such as permittivity, permeability, and conductivity, which collectively influence wave behavior. Its components depend on the wavevector (\(\kappa_n\)) and frequency (\(\omega_n\)), creating a direct relationship between the two. By enforcing \(\det{[\textbf{U}]} = 0\), the propagation criteria are established, revealing how frequency and wavevector are interconnected. This relationship, expressed as \(\omega_n(\kappa_n)\) or \(\kappa_n(\omega_n)\), provides insight into the system’s allowed frequencies, wavevectors, group velocity, phase velocity, and band structure. Thus, Eq.~\eqref{eqa:det_gen} serves as a critical criterion for wave propagation, and solving it yields the dispersion relation that defines wave dynamics within the medium.

The matrix \([\textbf{U}]\), along with the truncated set of \(g_n\) coefficients, illustrates the effects of nonlinearity. In nonlinear systems, energy from the primary harmonic modes (\(U_{nn}\)) is redistributed across an infinite series of other harmonic modes (\(U_{nm}\), where \(n \neq m\), \(-\infty < n, m < +\infty\)). This contrasts with linear space-time-modulated media~\cite{Taravati_PRAp_2018}, where energy is redistributed only among the nearest neighboring harmonic modes (\(U_{n,n-1}\) and \(U_{n,n+1}\)). Nonlinear interactions enable energy transfer across a wider range of harmonic frequencies, allowing the incident wave to excite higher-order harmonics efficiently. This is due to the ability of nonlinear systems to sustain the high-frequency modulation necessary for such energy redistribution.

\section{Wave Incidence and Transmission}\label{sec:inc}
The TM incident fields read
	\begin{subequations}
	\begin{equation}
\mathbf{H}_\text{I} (x,z,t)= \mathbf{\hat{y}} H_0 e^{(j \omega _0 t)}\cdot e^{-j\left[k_x x +k_0 \sin(\theta_\text{i}) z \right]}
	\end{equation}
\begin{equation}
	\begin{split}
&\mathbf{E}_\text{I} (x,z,t)= -\eta_1 \left[\mathbf{\hat{k}}_\text{I} \times \mathbf{H}_\text{I} (x,z,t)\right] \\&= \eta_1 \left[\mathbf{\hat{x}} \sin(\theta_\text{i}) -\mathbf{\hat{z}} \cos(\theta_\text{i}) \right]   \cdot  H_0 e^{j \omega _0 t}\cdot e^{-j\left[k_x x +k_0 \sin(\theta_\text{i}) z  \right]}
\end{split}
\end{equation}
	\end{subequations}
\noindent where $\eta_1=\sqrt{\mu_0 \mu_\text{1}/(\epsilon_0\epsilon_1)}$.  The reflected and transmitted electric fields outside of the slab may be defined as
\begin{subequations}\label{eqa:A-ER_ET_forw}
	\begin{equation}
		\mathbf{H}_\text{R} (x,z,t)= \mathbf{\hat{y}} \sum\limits_{n =  - \infty }^\infty  R_{n} e^{-j \left[ k_x x -k_{0n} \sin(\theta_n^\text{R}) z  -\omega_n t\right] } ,
		\label{eqa:A-E_refl_forw}
	\end{equation}
	\begin{equation}
				\begin{split}
	&	\mathbf{E}_\text{R} (x,z,t)= -\eta_1 [\mathbf{\hat{k}}_\text{R}  \times \mathbf{H}_\text{R} (x,z,t)]\\ &=-\eta_1 \left[\mathbf{\hat{x}} \sin(\theta_n^\text{R}) +\mathbf{\hat{z}} \cos(\theta_\text{i}) \right]   \cdot R_{n}   e^{-j \left[ k_x x -k_{0n} \sin(\theta_n^\text{R}) z  -\omega_n t\right] } ,
		\label{eqa:A-H_refl_forw}
			\end{split}
	\end{equation}
	\begin{equation}
		\mathbf{H}_\text{T} (x,z,t)= \mathbf{\hat{y}} \sum\limits_n    T_{n} e^{-j \left[ k_x x +k_{0n} \sin(\theta_n^\text{T}) z   -\omega_n t \right] } ,
		\label{eqa:A-E_trans_forw}
	\end{equation}
	\begin{equation}
		\begin{split}
		\mathbf{E}_\text{T} (x,z,t)&= -\eta_3 [\mathbf{\hat{k}}_\text{T}  \times \mathbf{H}_\text{T} (x,z,t)] \\&= \eta_3 \sum\limits_n  \left[\mathbf{\hat{x}} \sin(\theta_n^\text{T}) -\mathbf{\hat{z}} \cos(\theta_\text{i}) \right]   \\& \cdot T_{n} e^{-j \left[ k_x x +k_{0n} \sin(\theta_n^\text{T}) z -\omega_n t \right]} ,
		\label{eqa:A-H_trans_forw}
			\end{split}
	\end{equation}
\end{subequations}
\noindent where $\eta_3=\sqrt{\mu_0\mu_\text{3}/(\epsilon_0\epsilon_3)}$.
continuity of the tangential components of the electromagnetic fields at $z=0$ and $z=d$ to find the unknown field amplitudes $H_{0}$, $R_{n}$ and $T_{n}$. The electric and magnetic fields continuity conditions between regions 1 and 2 at $z=0$, ${H_{1y}}(x,0,t) = {H_{2y}}(x,0,t)$ and ${E_{1x}}(x,0,t) = {E_{2x}}(x,0,t)$ read
\begin{subequations}
\begin{equation}\label{eqa:EBC_forw_12}
	\delta_{n0} H_0  + R_{n} =    H_{n},
\end{equation}
\begin{equation}\label{eqa:HBC_forw_12}
	\begin{split}
&	\eta_1 \sin(\theta_\text{i})   \delta_{n0} H_0   - \eta_1 \sin(\theta_n^\text{R}) R_{n}   =  \eta_2  \sin(\theta_{n}) H_{n}.
	\end{split}
\end{equation}

By solving the above two equations simultaneously, we achieve
\begin{equation}\label{eqa:aaa}
H_{0}=H_0   \dfrac{2\eta_1 \sin(\theta_\text{i})} {\eta_1 \sin(\theta_\text{i})  +\eta_2 \sin(\theta_{0})  } ,
\end{equation}

\begin{equation}\label{eqa:bbb}
	 R_{n} =       H_{n}-\delta_{n0} H_0 .
\end{equation}
\end{subequations}

The continuity condition of electric and magnetic fields between regions 2 and 3 at $z=d$, ${H_{2y}}(x,d,t) = {H_{3y}}(x,d,t)$ and ${E_{2x}}(x,d,t) = {E_{3x}}(x,d,t)$, reduces to
\begin{subequations}
\begin{equation}
 H_{n} e^{-j \kappa_{n} d}  = T_{n}  e^{- j k_{0n} \sin(\theta_n^\text{T}) d},
	\label{eqa:EBC_forw_23}
\end{equation}
\begin{equation}
	\begin{split}
	\eta_2	  &  \sin(\theta_n)  H_{n}  e^{-j \kappa_{n} d} = \eta_3 \sin(\theta_n^\text{T}) T_{n} e^{- j k_{n} \sin(\theta_n^\text{T}) d},
	\end{split}
	\label{eqa:HBC_forw_23}
\end{equation}
which gives
\begin{equation}
	T_{n}  =
 \dfrac{\eta_2	    \sin(\theta_n)}{\eta_3 \sin(\theta_n^\text{T})} H_{n} e^{-j [\kappa_{n}- k_{n} \sin(\theta_n^\text{T})] d}.
	\label{eqa:Tn}
\end{equation}
\end{subequations}

The scattering angles of the transmitted space-time harmonics, denoted as $\theta_n^{\text{T}}$, can be derived from the Helmholtz relations. These relations ensure that the wavevector components satisfy the dispersion constraints within the space-time-modulated medium. Mathematically, this is expressed as
\begin{subequations} 
\begin{equation}\label{eqa:angl}
 [k_{0} \cos(\theta_{\text{i}})]^2 + [k_{n} \sin(\theta_n^\text{T})]^2 = k_{n}^2, 
 \end{equation} 
  where $k_0$ is the wavevector magnitude of the incident wave, $\theta_\text{i}$ is the angle of incidence, $k_n$ is the wavevector magnitude of the transmitted $n$-th harmonic, and $\theta_n^\text{T}$ is the corresponding angle of transmission. By solving for $\theta_n^\text{T}$, the following relationship is obtained
	\begin{equation}\label{eqa:trans_angl} 
		\theta_n^{\text{T}} = \sin^{-1} \left(\frac{\cos(\theta_\text{i})}{1 + n\omega_\text{s} / \omega_0} \right),
	 \end{equation} 
 \end{subequations} 
 where $\omega_0$ is the frequency of the incident wave, $\omega_\text{s}$ is the space-time modulation frequency, and $n$ is the harmonic index, which takes integer values representing the order of the generated harmonics.

Equation~\eqref{eqa:angl} reflects the conservation of the transverse component of the wavevector at the interface, accounting for the interaction between the incident wave and the space-time modulation. The modulation imparts additional energy and momentum to the wave, resulting in the generation of harmonics with modified frequencies and wavevectors. The resulting transmission angle $\theta_n^\text{T}$, given by Equation~\eqref{eqa:trans_angl}, depends on both the incident angle $\theta_\text{i}$ and the modulation frequency ratio $\omega_\text{s} / \omega_0$. This indicates that the scattering angles of the transmitted harmonics are not only influenced by the geometry of incidence but also by the properties of the modulation, such as its frequency and strength. Physically, this relationship highlights the angular dispersion of the harmonics, where higher-order harmonics ($n > 0$) are transmitted at increasingly steeper angles, while lower-order or subharmonics may exhibit overlapping or shallow angles of transmission. The tunability of $\omega_\text{s}$ provides a mechanism to control the angular and frequency distribution of the transmitted waves, offering potential applications in beam steering, frequency-angular multiplexing, and dynamic wavefront shaping. This analysis underscores the utility of space-time-modulated media in advanced wave manipulation, enabling functionalities that are unattainable with conventional static structures.

\section{Frequency-Angular Beam Multiplexing}\label{sec:exp}
\begin{figure}
	\begin{center}
		\includegraphics[width=1\columnwidth]{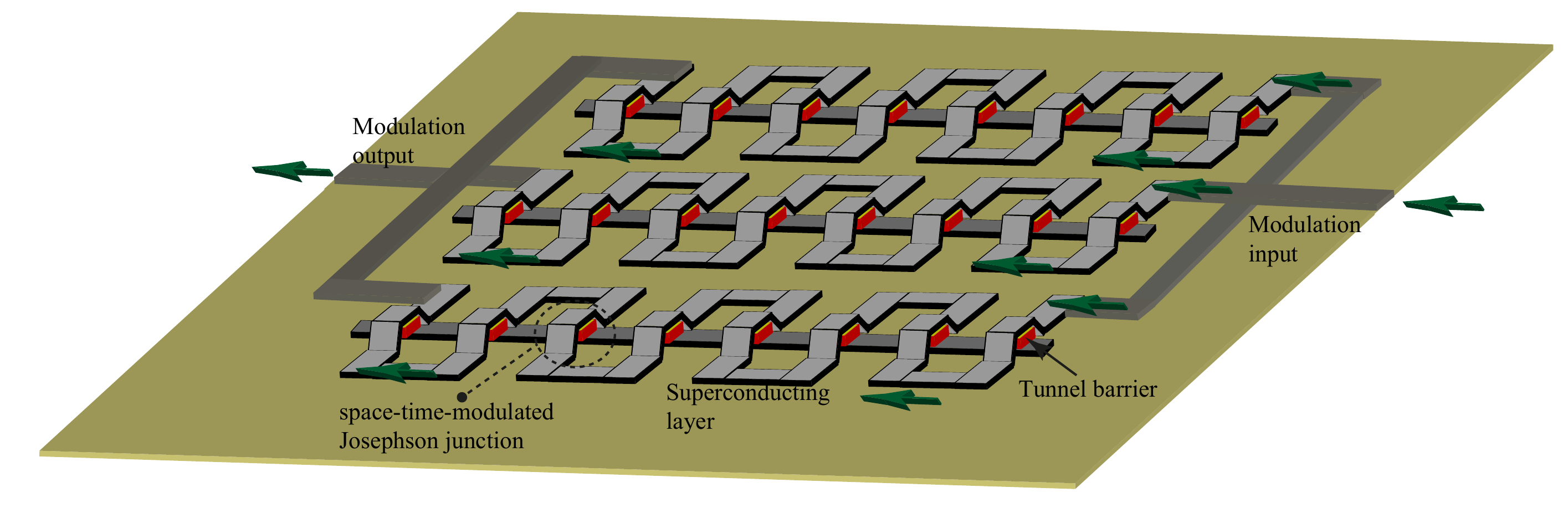}
		\caption{Experimental prototype design of a space-time-modulated Josephson junction array.}
		\label{Fig:exp}
	\end{center}
\end{figure}

Figure~\ref{Fig:exp} illustrates the experimental prototype design of a space-time-modulated Josephson junction array. Each Josephson junction is formed by a thin dielectric tunnel barrier sandwiched between two superconducting layers. The superconducting layers are typically fabricated using materials such as aluminum (Al) or niobium (Nb), chosen for their high critical current densities, low microwave losses, and compatibility with cryogenic environments essential for quantum applications. The Josephson junctions are patterned on a high-quality substrate, commonly silicon (Si) or sapphire (Al\(_2\)O\(_3\)), both known for their excellent thermal conductivity and low dielectric loss at cryogenic temperatures. The substrate ensures mechanical stability and minimizes parasitic effects, such as crosstalk and substrate-induced losses, which are critical for high-frequency and high-fidelity operations. To achieve space-time modulation, a modulation signal is applied to the Josephson junctions through a power divider on the right-hand side of the setup. This power divider ensures the uniform distribution of the modulation signal across the array, enabling precise control over the space-time-dependent electromagnetic properties. The modulation signal is typically delivered through high-quality coaxial cables or superconducting transmission lines to minimize attenuation and phase noise. On the left side, another power divider directs the modulation signal to a matched load, ensuring minimal signal reflection. This matched load is carefully chosen to provide impedance matching across the modulation frequency range, reducing standing waves and preserving the integrity of the modulation waveform. The entire setup is housed within a cryogenic environment, often a dilution refrigerator, to maintain the superconducting state of the layers and to suppress thermal noise. The cryogenic setup also incorporates electromagnetic shielding to minimize external interference and ensure the stability of the space-time modulation. The precise engineering of the substrate, materials, and modulation delivery plays a pivotal role in achieving the desired performance and reproducibility.

\begin{figure}
	\begin{center}
		\includegraphics[width=0.8\columnwidth]{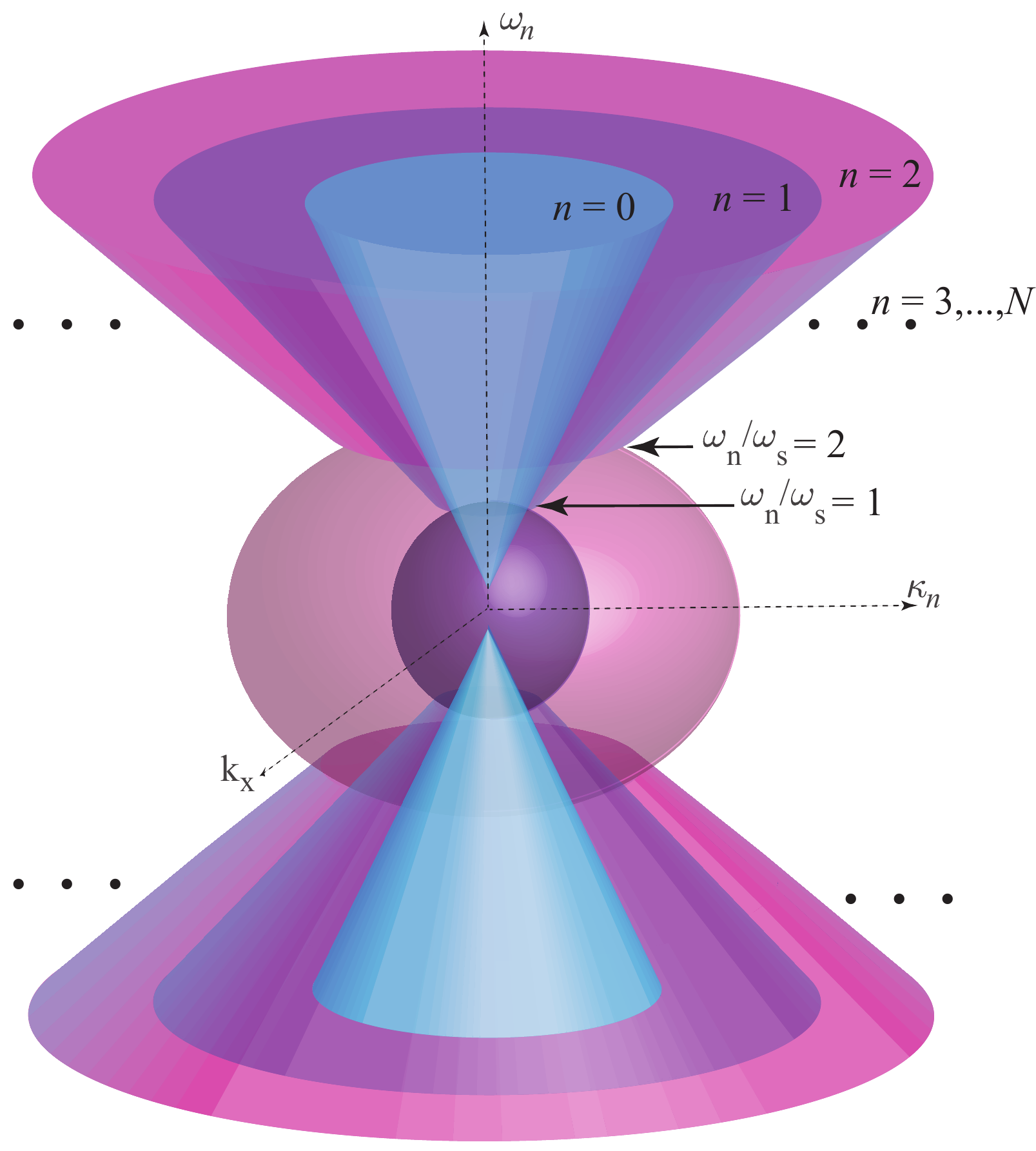}
		\caption{Three-dimensional dispersion diagram of space-time-periodic Josephson junctions, illustrating the relationship between frequency $\omega_n$, the $z$-component of the wavevector $\kappa_n$, the $x$-component of the wavevector $k_x$ representing the angle of incidence $\theta_\text{i}=\cos^{-1}(k_x/k_0)$, and modulation frequency $\omega_\text{s}$.}
			\label{Fig:3Ddisp}
		\end{center}
\end{figure}

According to the dispersion relation in Eq.~\eqref{eqa:det_gen}, a three-dimensional dispersion diagram can be constructed to visualize the wave dynamics in space-time-periodic Josephson junctions. Figure~\ref{Fig:3Ddisp} qualitatively depicts this three-dimensional dispersion diagram, illustrating the interdependence of the frequency (\(\omega_n\)), the \(z\)-component of the wavevector (\(\kappa_n\)), the \(x\)-component of the wavevector (\(k_x\)), and the modulation frequency (\(\omega_\text{s}\)). The \(x\)-component of the wavevector, \(k_x\), is particularly significant as it represents the angle of incidence (\(\theta_\text{i} = \cos^{-1}(k_x / k_0)\)). The dispersion diagram features a series of double semi-cones aligned along the \(\omega_n\) axis, each corresponding to a specific space-time harmonic. The positive-slope regions of the semi-cones, with respect to the \((\omega_n, \kappa_n)\) plane, represent a positive group delay corresponding to forward-propagating waves, while the negative-slope regions indicate negative group delay corresponding to backward-propagating waves. The steepness of the semi-cones is determined by the space-time modulation parameter, \(\widetilde{\Phi}_\text{dc}\), such that increasing \(\widetilde{\Phi}_\text{dc}\) results in a greater slope, reflecting enhanced modulation strength. At the center of each double semi-cone lies an electromagnetic bandgap, which is influenced by \(\widetilde{\Phi}_\text{dc}\).As \(\widetilde{\Phi}_\text{dc}\) increases, the bandgap widens, signifying enhanced coupling between harmonics and more efficient energy redistribution across them. The slope of the higher-order space-time harmonic cones is directly proportional to the space-time modulation amplitude \(\widetilde{\Phi}_\text{rf}\). When \(\widetilde{\Phi}_\text{rf} \to 0\), the higher-order harmonic cones vanish entirely, reflecting the absence of dynamic modulation and reducing the system to a purely static problem. This intricate interplay between the modulation parameters, \(\widetilde{\Phi}_\text{dc}\) and \(\widetilde{\Phi}_\text{rf}\), highlights the tunability of the system’s dispersion characteristics and provides valuable insights into the propagation and manipulation of waves in space-time-modulated Josephson junctions.
	
\begin{figure*}
	\begin{center}
		\subfigure[]{\label{Fig:disp1}
			\includegraphics[width=0.31\linewidth]{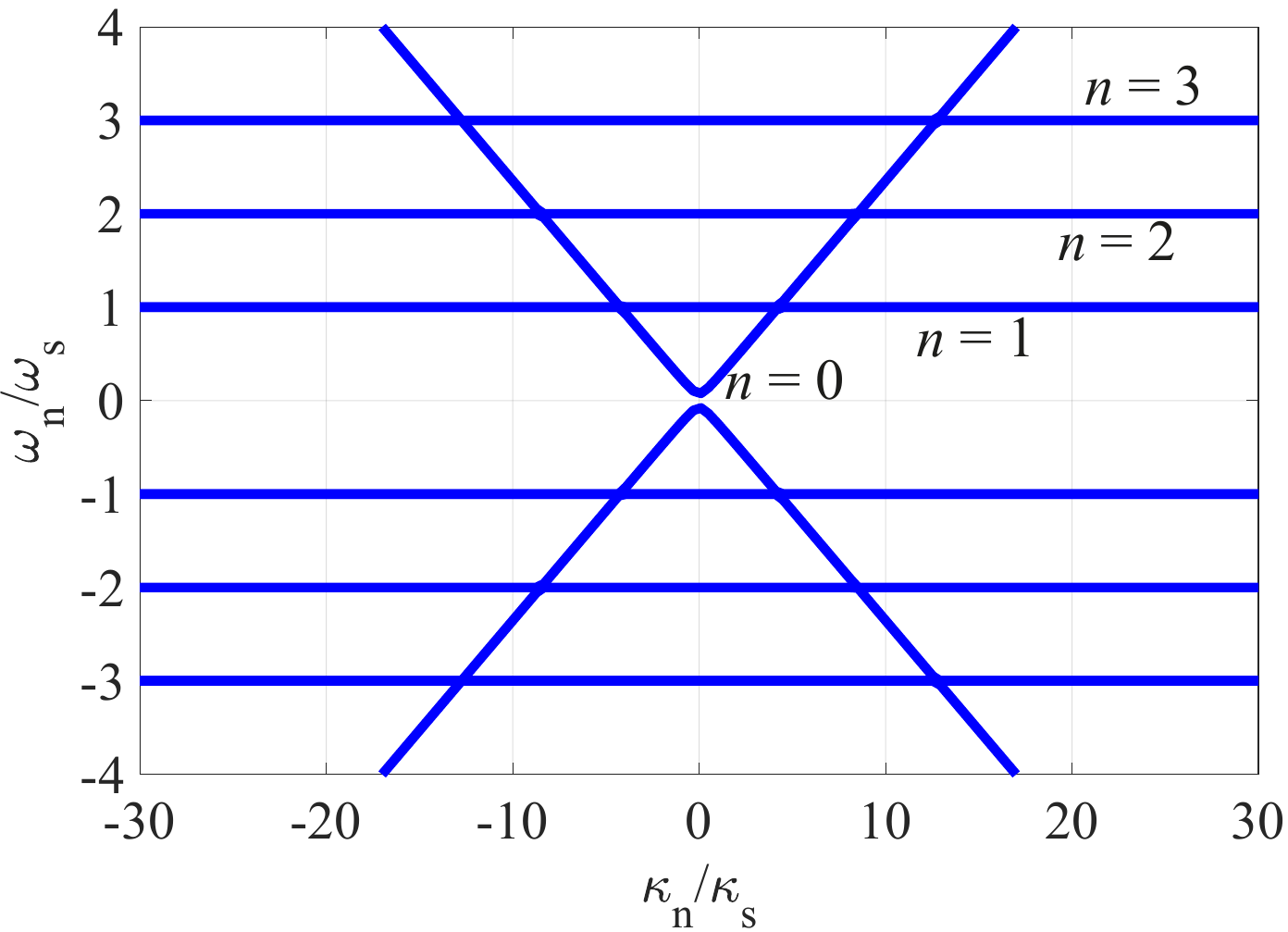}  }
		\subfigure[]{\label{Fig:disp2}
			\includegraphics[width=0.31\linewidth]{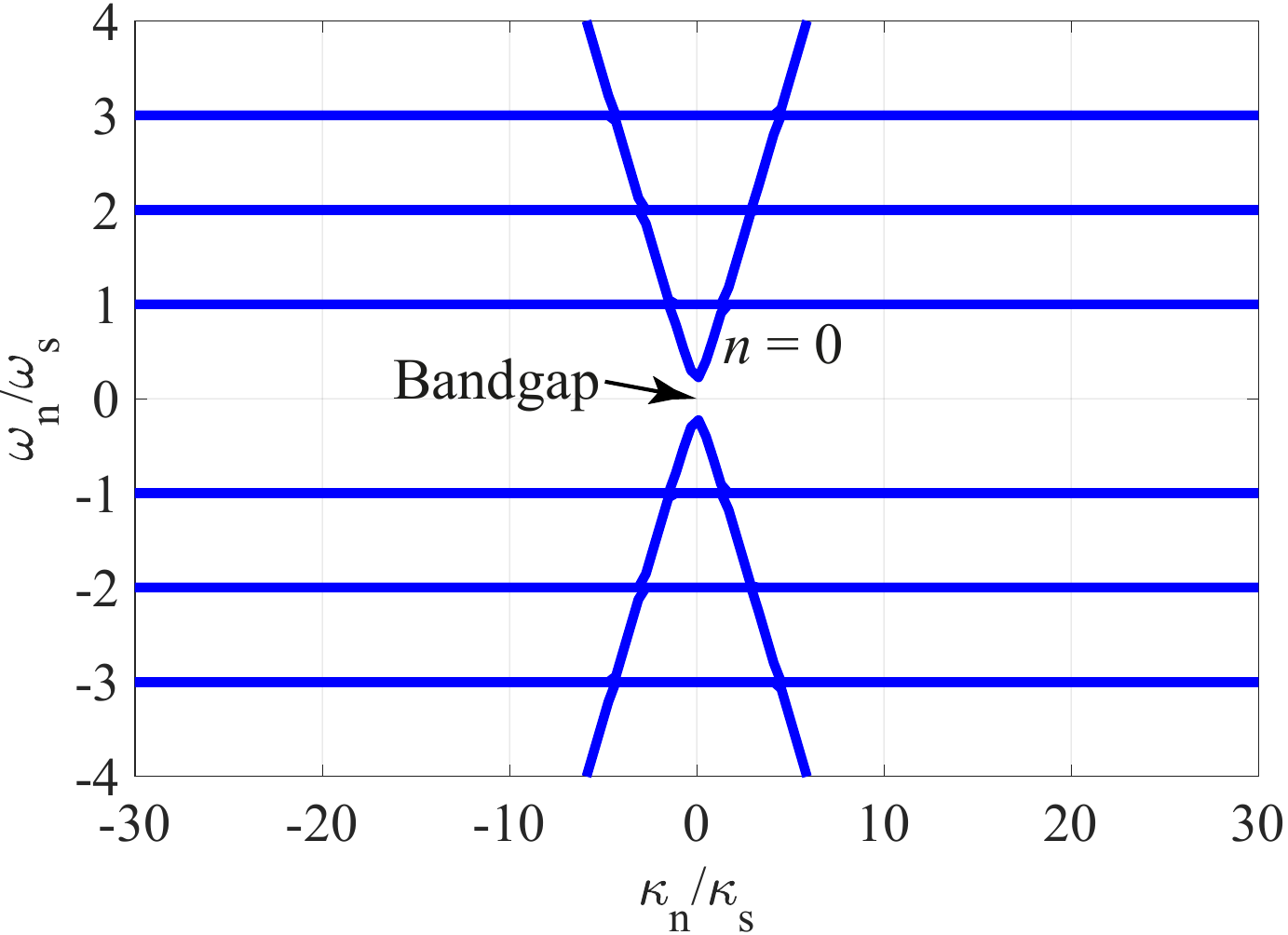}  }
		\subfigure[]{\label{Fig:disp3}
			\includegraphics[width=0.31\linewidth]{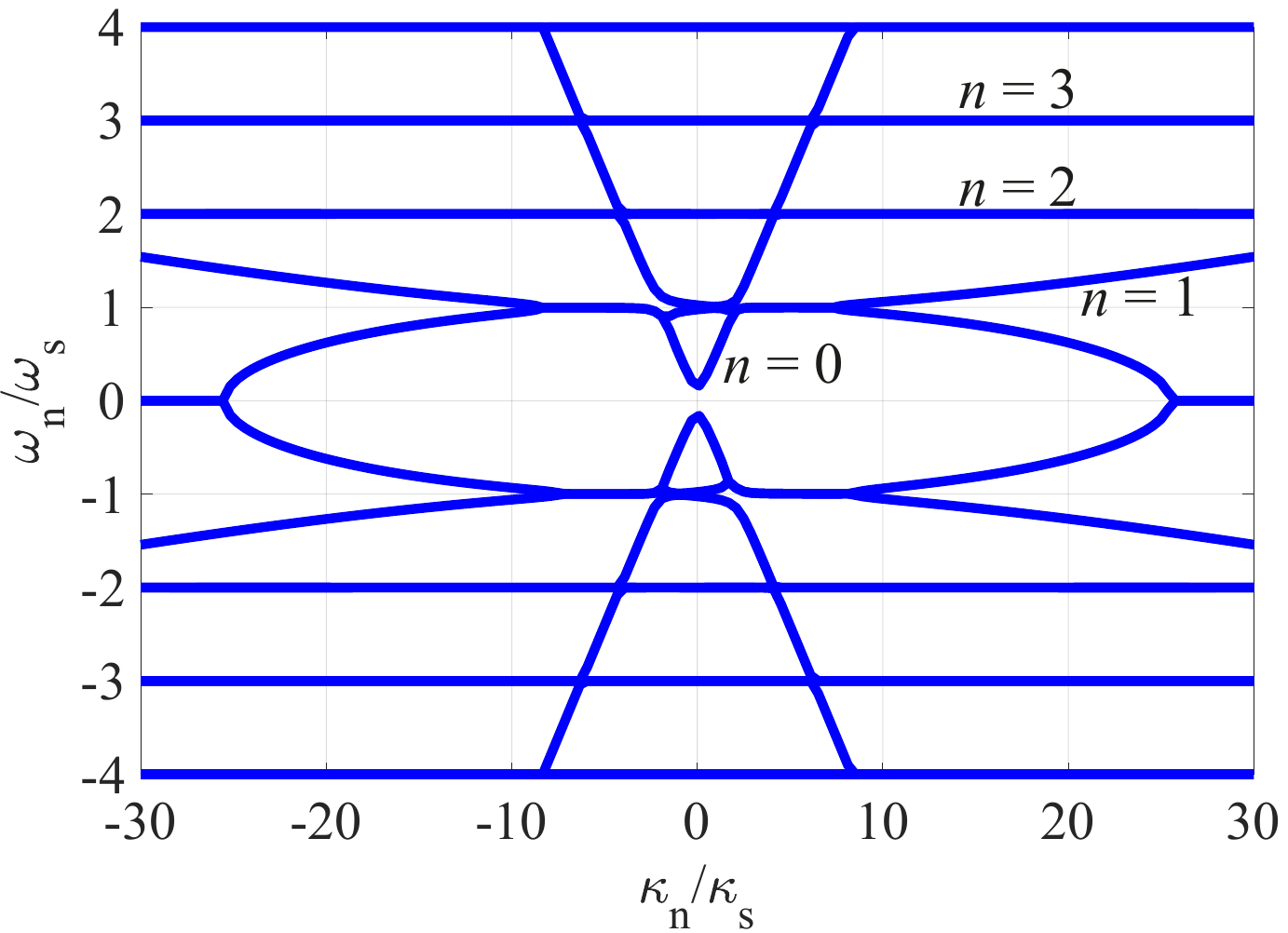}  }
		\subfigure[]{\label{Fig:disp4}
			\includegraphics[width=0.31\linewidth]{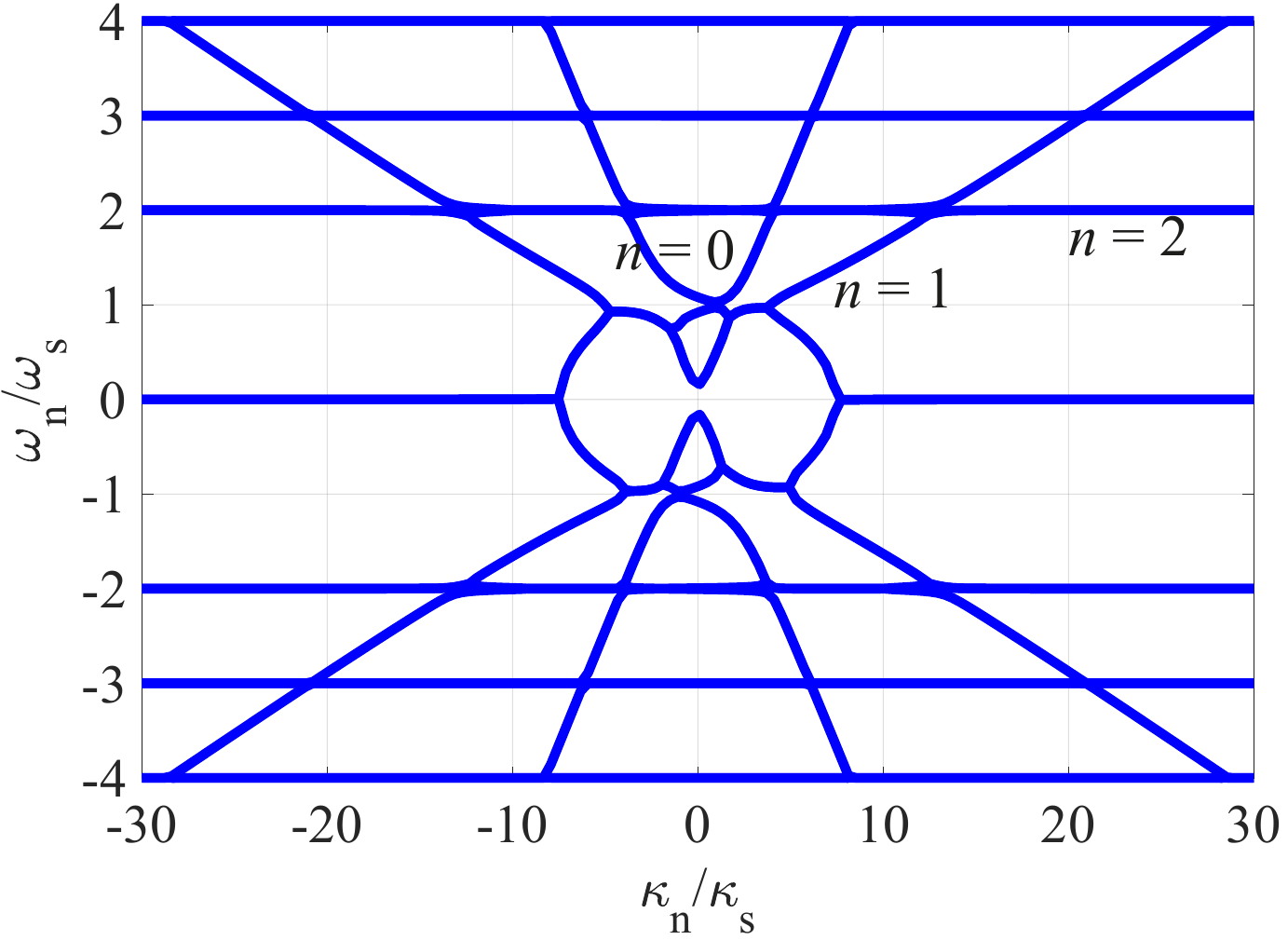}  }
		\subfigure[]{\label{Fig:disp5}
			\includegraphics[width=0.31\linewidth]{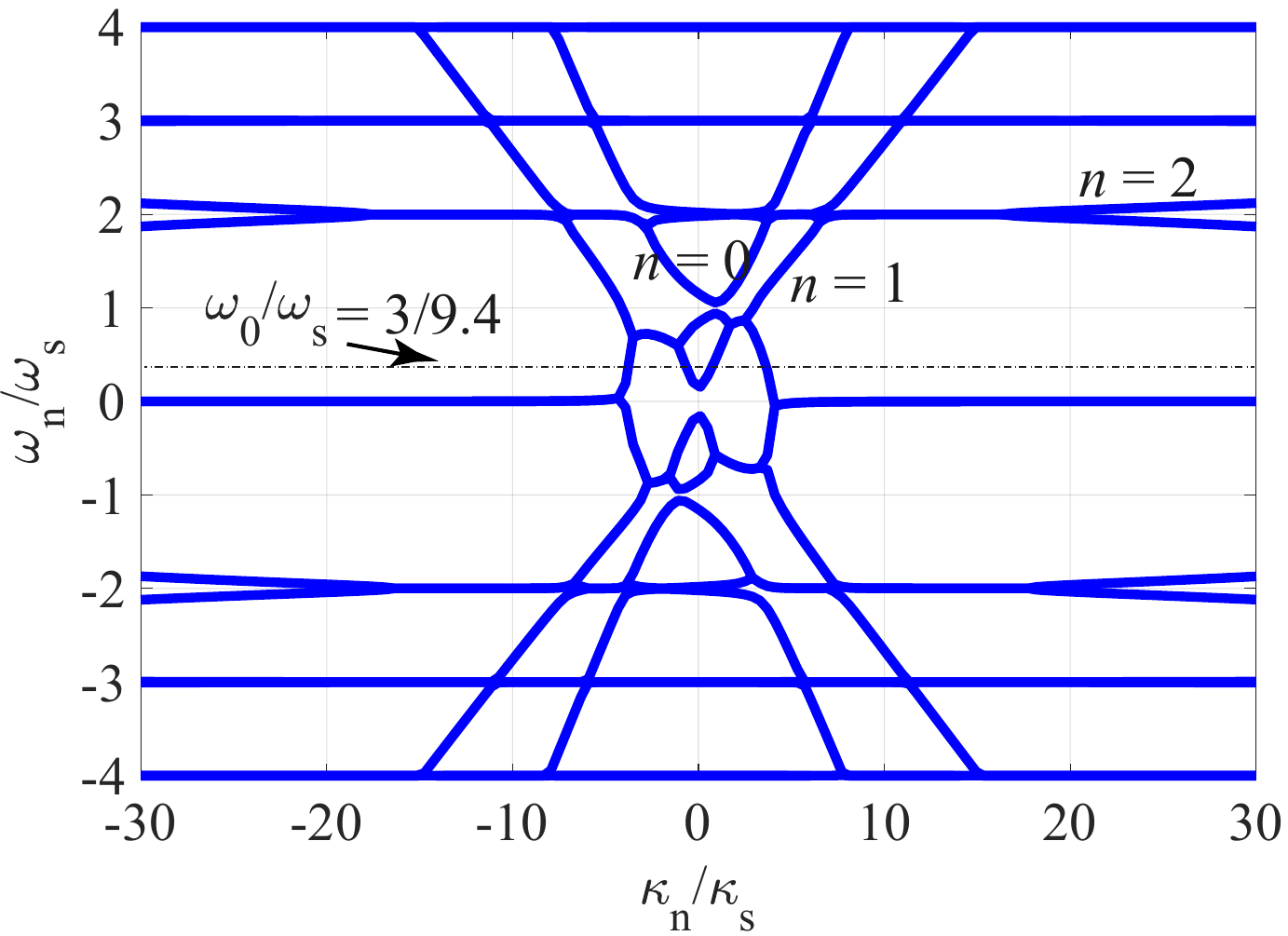}  }
		\subfigure[]{\label{Fig:disp6}
			\includegraphics[width=0.31\linewidth]{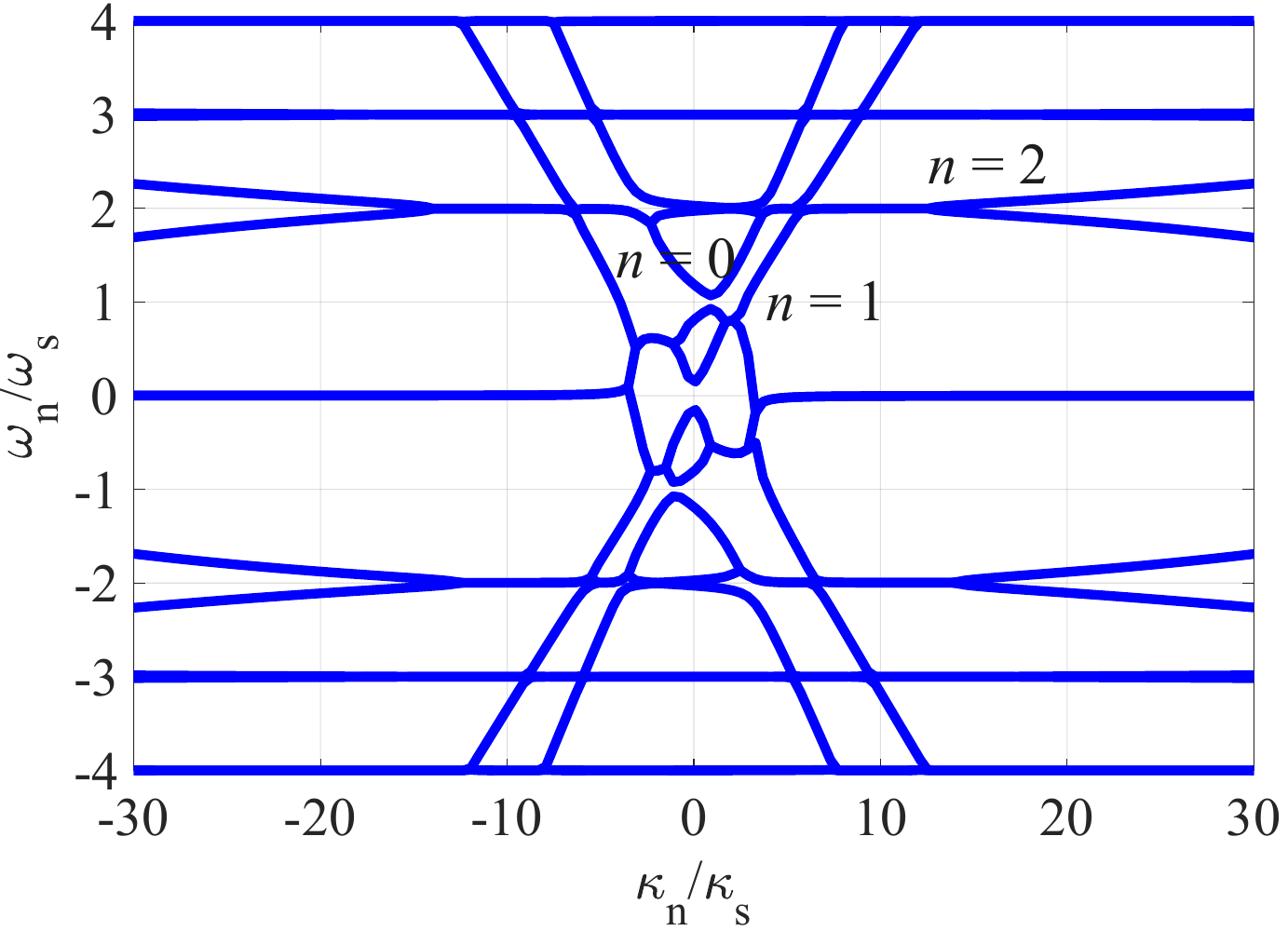}  }
		\caption{Dispersion diagrams for different operation regimes of the space-time-modulated Josephson junction array. (a)~$\widetilde{\Phi}_\text{dc} \to 0$ and $\widetilde{\Phi}_\text{rf} \to 0$. (b)~$\widetilde{\Phi}_\text{dc} =0.7$ and $\widetilde{\Phi}_\text{rf} \to 0$. (c)~$\widetilde{\Phi}_\text{dc} =0.7$ and $\widetilde{\Phi}_\text{rf} =0.1$. (d)~$\widetilde{\Phi}_\text{dc} =0.7$ and $\widetilde{\Phi}_\text{rf} =0.35$. (e)~$\widetilde{\Phi}_\text{dc} =0.7$ and $\widetilde{\Phi}_\text{rf} =0.7$. (f)~$\widetilde{\Phi}_\text{dc} =0.7$ and $\widetilde{\Phi}_\text{rf} =0.9$.}
		\label{Fig:disp}
	\end{center}
\end{figure*}

Figures~\ref{Fig:disp1} through \ref{Fig:disp6} present the analytical two-dimensional dispersion diagrams (\(\omega_n\) versus \(\kappa_n\)) for various operational regimes of the space-time-modulated Josephson junction array. These diagrams are calculated using Eq.~\eqref{eqa:det_gen}, illustrating the influence of the space-time modulation parameters, \(\widetilde{\Phi}_\text{dc}\) (static modulation) and \(\widetilde{\Phi}_\text{rf}\) (dynamic modulation), on the system's dispersion characteristics. Figure~\ref{Fig:disp1} illustrates the case where both modulation parameters are negligible (\(\widetilde{\Phi}_\text{dc} \to 0\) and \(\widetilde{\Phi}_\text{rf} \to 0\)), resulting in a conventional dispersion diagram. In this configuration, no bandgaps or harmonic coupling are observed, and higher-order space-time harmonic cones (\(n > 0\)) are absent, indicating the lack of significant space-time modulation effects. Figure~\ref{Fig:disp2} shows the scenario where \(\widetilde{\Phi}_\text{dc} = 0.7\) and \(\widetilde{\Phi}_\text{rf} \to 0\). In this case, the static modulation causes the fundamental harmonic (\(n = 0\)) dispersion curve to tilt. Additionally, a bandgap appears at the center of the \(n = 0\) double semi-cone, highlighting the influence of static modulation on the dispersion properties. Figure~\ref{Fig:disp3} demonstrates the effect of a small dynamic modulation amplitude (\(\widetilde{\Phi}_\text{rf} = 0.1\)) superimposed on the static modulation (\(\widetilde{\Phi}_\text{dc} = 0.7\)). This configuration weakly excites the $n=1$ harmonic and introduces narrow electromagnetic bandgaps at the harmonic intersections, signifying weak coupling between the harmonics. Figure~\ref{Fig:disp4} illustrates the dispersion for \(\widetilde{\Phi}_\text{dc} = 0.7\) and \(\widetilde{\Phi}_\text{rf} = 0.35\). With the increased dynamic modulation amplitude, the slope of the \(n = 1\) harmonic becomes steeper, indicating stronger coupling between the \(n = 0\) and \(n = 1\) harmonics. This results in more significant energy redistribution between these harmonics, enhancing the interaction and modulation effects within the system. 

Figure~\ref{Fig:disp5} illustrates the case where \(\widetilde{\Phi}_\text{dc} = 0.7\) and \(\widetilde{\Phi}_\text{rf} = 0.7\). In this scenario, the slope of the \(n = 1\) harmonic increases noticeably, reflecting stronger space-time modulation effects. Additionally, higher-order harmonics, including the \(n = 2\) harmonic, are weakly excited, indicating the emergence of more complex wave interactions. The dispersion curves reveal significant coupling between the \(n = 0\) and \(n = 1\) harmonics, highlighting enhanced energy transfer and interaction due to the increased modulation amplitude. This regime highlights the system's ability to manipulate wave propagation robustly. We emphasize the asymmetric coupling between the forward and backward space-time harmonics, which gives rise to nonreciprocal wave propagation in these arrays. This nonreciprocal behavior opens up new possibilities for a wide range of applications, similar to those seen in conventional linear space-time-modulated arrays, but with the added advantage of directional control over wave propagation~\cite{Taravati_Kishk_MicMag_2019,Taravati_ACSP_2022}. Figure~\ref{Fig:disp6} corresponds to a higher dynamic modulation amplitude considered (\(\widetilde{\Phi}_\text{rf} = 0.9\)), where the slope of the higher order harmonics is further increased, and the harmonics exhibit asymmetric coupling. This configuration underscores the impact of strong space-time modulation on wave behavior, enabling advanced functionalities such as efficient energy transfer to higher-order harmonics. These diagrams collectively demonstrate the tunable nature of the dispersion characteristics in space-time-modulated Josephson junction arrays. By adjusting the modulation parameters, one can control key wave propagation features, including harmonic coupling strength, nonreciprocity, and bandgap width, providing a versatile platform for wave manipulation in advanced quantum and electromagnetic systems.

\begin{figure*}
	\begin{center}
		\subfigure[]{\label{Fig:iso1}
			\includegraphics[width=0.31\linewidth]{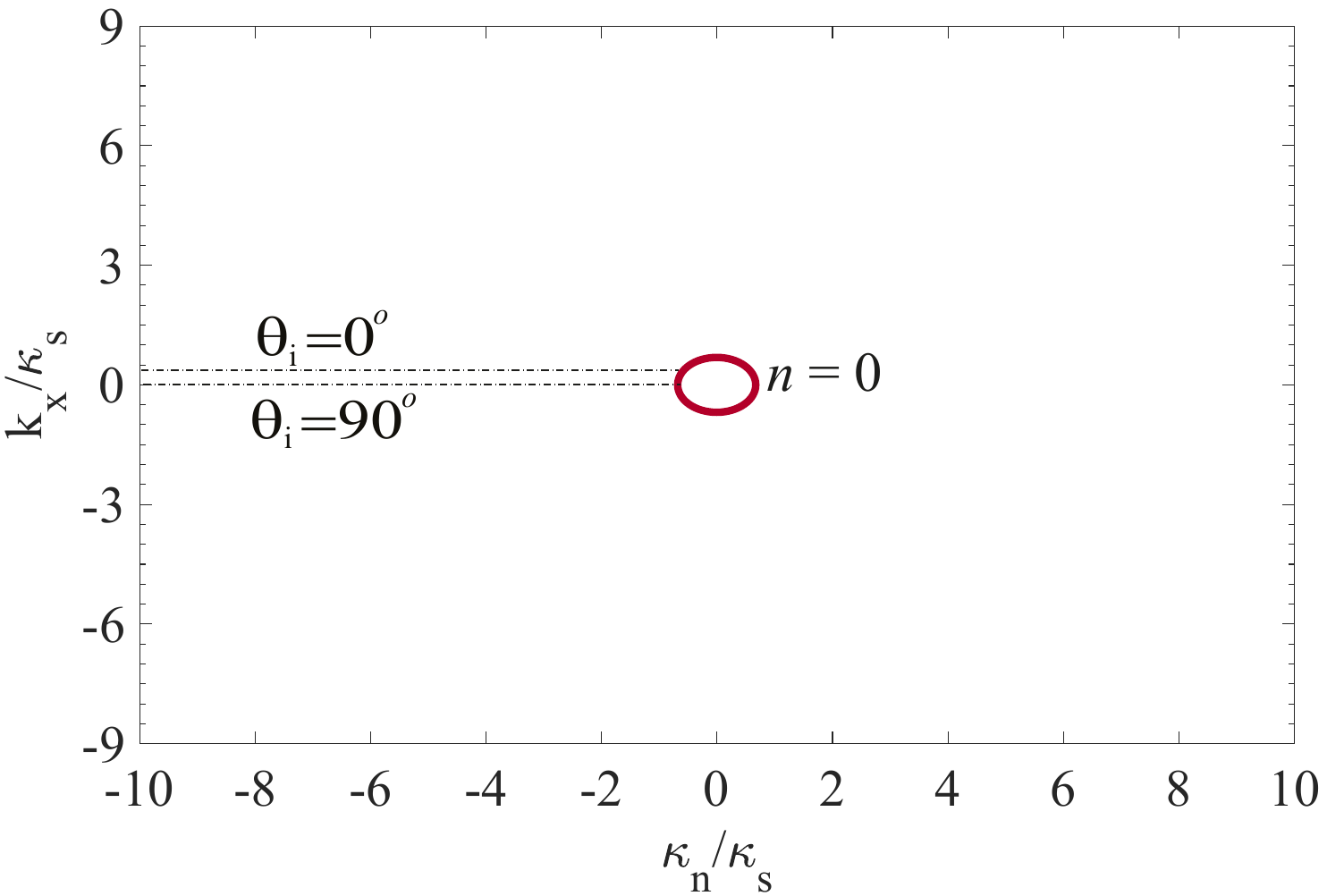}  }
		\subfigure[]{\label{Fig:iso2}
			\includegraphics[width=0.31\linewidth]{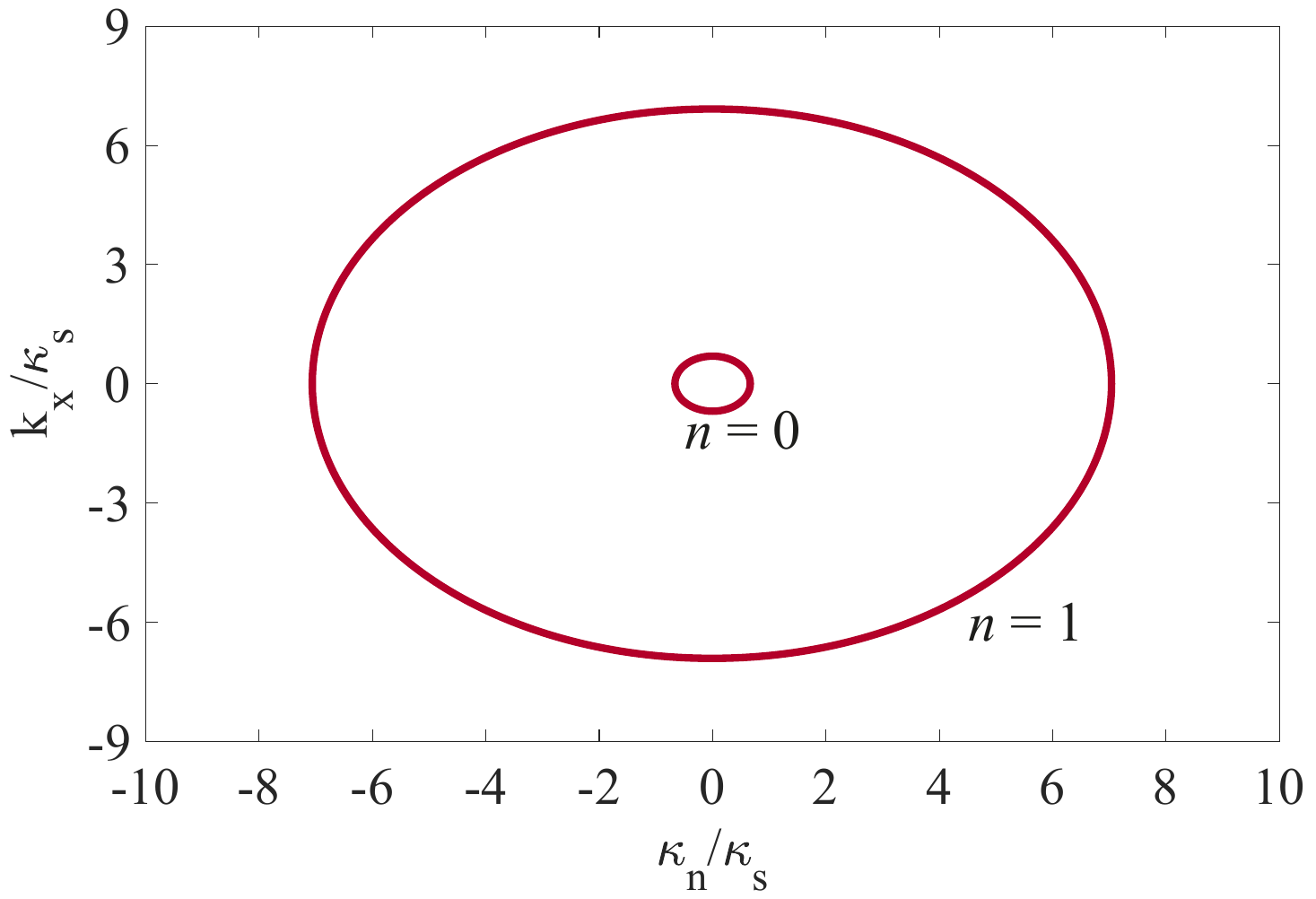}  }
		\subfigure[]{\label{Fig:iso3}
			\includegraphics[width=0.31\linewidth]{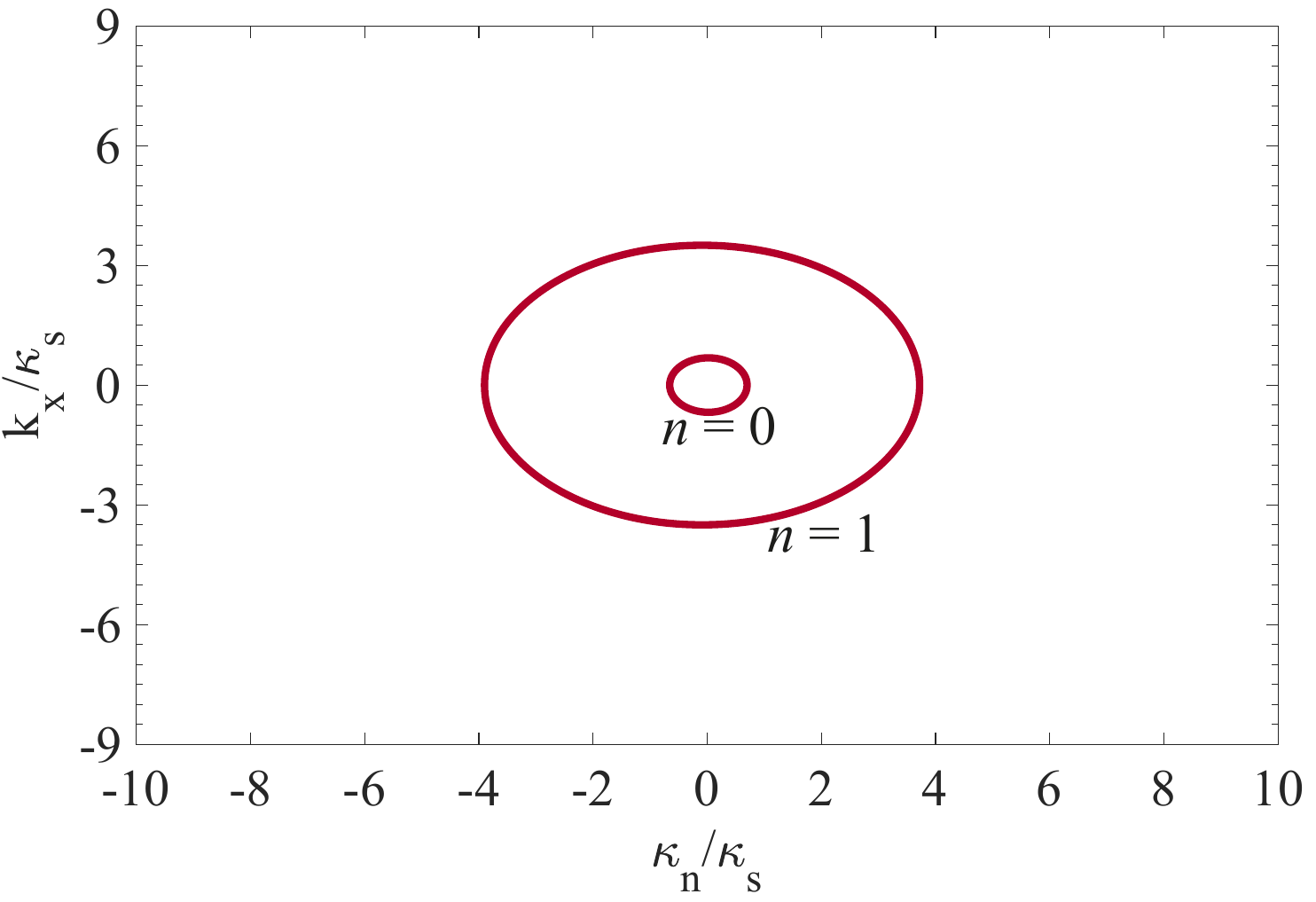}  }
		\caption{Isofrequency diagrams for different operation regimes of the space-time-modulated Josephson junction array at $\omega_0/\omega_\text{s}=3/9.4$. (a)~$\widetilde{\Phi}_\text{dc} =0.7$ and $\widetilde{\Phi}_\text{rf} \to 0$. (b)~$\widetilde{\Phi}_\text{dc} =0.7$ and $\widetilde{\Phi}_\text{rf} =0.35$. (c)~$\widetilde{\Phi}_\text{dc} =0.7$ and $\widetilde{\Phi}_\text{rf} =0.7$.}
		\label{Fig:iso}
	\end{center}
\end{figure*}

Figures~\ref{Fig:iso1} to ~\ref{Fig:iso3} present the analytical isofrequency diagrams for different operational regimes of the space-time-modulated Josephson junction array, computed using Eq.~\eqref{eqa:det_gen} at $\omega_0/\omega_\text{s}=3/9.4$. These diagrams provide valuable insight into the relationship between the wavevector components and the frequency at different modulation amplitudes. In Figure~\ref{Fig:iso1}, where \(\widetilde{\Phi}_\text{dc} = 0.7\) and \(\widetilde{\Phi}_\text{rf} \to 0\), we observe the behavior of the system under static modulation. The isofrequency $n=0$ contour displays a conventional shape, without any noticeable deformation, as higher-order space-time harmonics are not excited. This reflects the lack of dynamic coupling between the harmonics, indicating that the system behaves in a largely linear manner, similar to traditional dispersion relationships in non-modulated systems. In Figure~\ref{Fig:iso2}, with \(\widetilde{\Phi}_\text{dc} = 0.7\) and \(\widetilde{\Phi}_\text{rf} = 0.35\), the system is subjected to dynamic modulation, leading to noticeable changes in the isofrequency contours. The contours exhibit a more pronounced curvature, reflecting stronger interactions between the fundamental and first harmonic modes. The system now demonstrates a higher degree of nonreciprocal behavior and enhanced coupling between the $n=0$ and $n=1$ harmonics, influencing wave propagation characteristics.

Figure~\ref{Fig:iso3}, where \(\widetilde{\Phi}_\text{dc} = 0.7\) and \(\widetilde{\Phi}_\text{rf} = 0.7\), shows a further increase in dynamic modulation, leading to even more pronounced alterations in the isofrequency diagrams. At this higher modulation amplitude, the contour for the \(n=1\) harmonic becomes much closer to the \(n=0\) contour, indicating a significantly stronger coupling and enhanced energy exchange between these two harmonics. This closer proximity reflects the substantial redistribution of energy between the fundamental and first harmonics, highlighting the increased interaction due to the dynamic modulation. The isofrequency contours become significantly distorted, illustrating strong interactions across a wider range of higher-order harmonics, further revealing the complexity introduced by the increased modulation. The deformations in the isofrequency contours suggest a departure from the conventional linear dispersion, emphasizing the importance of dynamic modulation in shaping wave propagation and harmonic coupling within the system. Overall, these isofrequency diagrams reveal how different modulation parameters—particularly \(\widetilde{\Phi}_\text{dc}\) and \(\widetilde{\Phi}_\text{rf}\)—affect the wave propagation characteristics and harmonic coupling in the space-time-modulated Josephson junction array.

\begin{figure}
	\begin{center}
		\includegraphics[width=1\columnwidth]{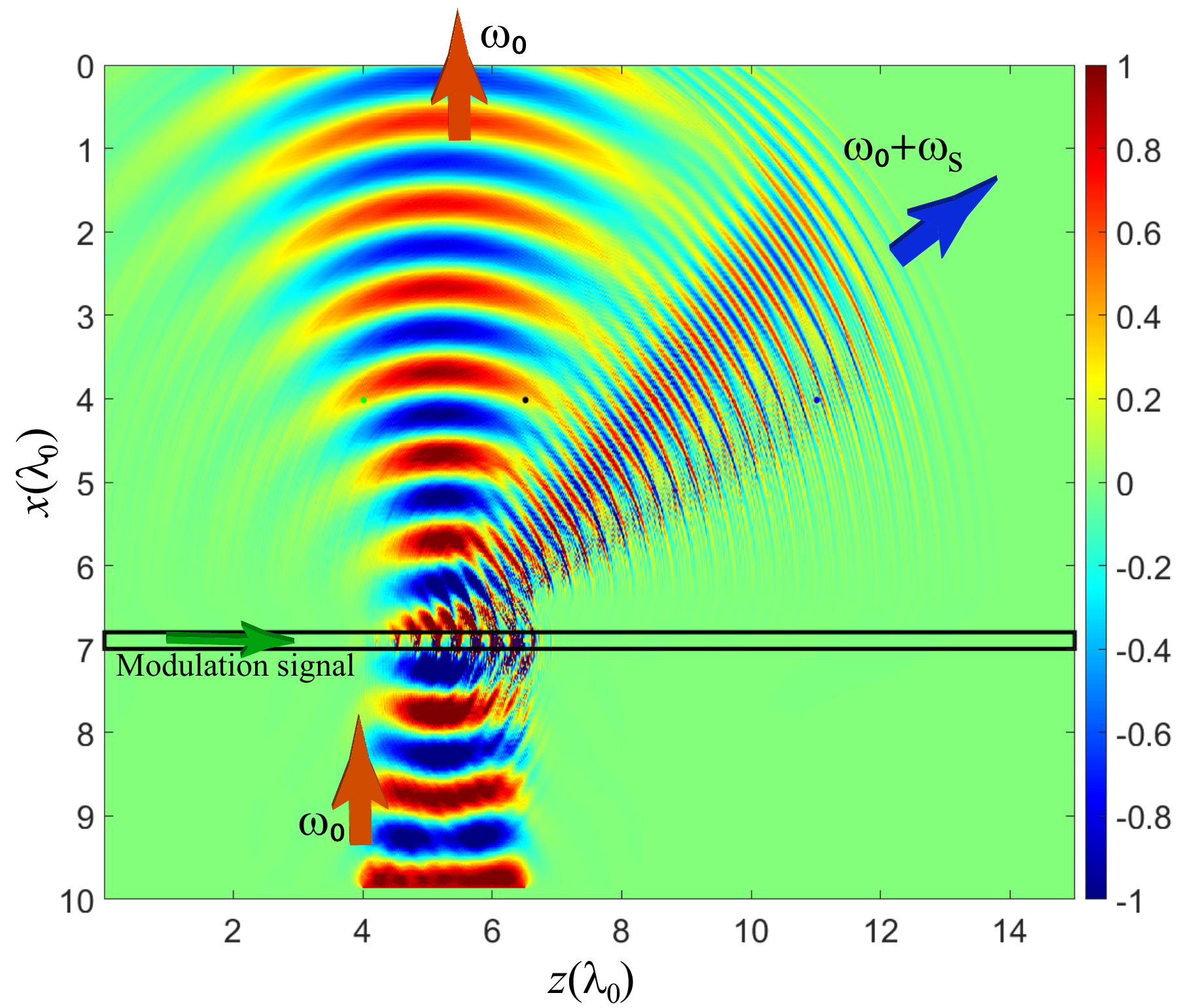}
		\caption{FDTD numerical simulation results for the $H_z$ field distribution demonstrate angular-frequency beam multiplexing, where the incident beam from the bottom is transmitted to the top, along with an up-converted beam transmitted at a different angle of transmission.}
		\label{Fig:res}
	\end{center}
\end{figure}

Figure~\ref{Fig:res} showcases the FDTD numerical simulation results depicting the $H_z$ field distribution. The simulation considers an incident wave with a frequency of $\omega_0 = 2\pi \times 3$~GHz, interacting with a space-time modulation characterized by a temporal modulation frequency of $\omega_\text{s} = 2\pi \times 9.4$~GHz. The modulation parameters are set as $\widetilde{\Phi}_\text{dc} = 0.7$ and $\widetilde{\Phi}_\text{rf} = 0.7$, representing the static and dynamic components of the modulation amplitude, respectively. The simulations demonstrate angular-frequency beam multiplexing, a phenomenon where an incident electromagnetic beam at frequency $\omega_0$, arriving from the bottom of the space-time-modulated structure, undergoes complex interactions within the medium. As a result, part of the beam is transmitted to the top at the same frequency $\omega_0$, following the principle of conventional wave transmission. In addition to this primary transmission, the modulated structure enables the generation of an up-converted beam. This up-converted beam emerges at a different angle of transmission compared to the incident beam and carries a distinct frequency, $\omega_0 + \omega_\text{s}=4.133\omega_0$. This frequency conversion is a hallmark of the space-time modulation, where the temporal and spatial variations in the material properties impart additional energy and momentum to the propagating wave. The angle of transmission for the up-converted beam is determined by the conservation of energy and momentum in the presence of the dynamic modulation. The interplay between the modulation frequency and the spatial periodicity of the modulated structure governs the angular dispersion and the efficiency of the frequency conversion process. These factors can be engineered by tailoring the modulation parameters, such as the modulation depth, frequency, and spatial profile, to achieve desired beam steering and frequency-multiplexing effects. Such a phenomenon has significant implications for advanced communication and sensing applications. By enabling simultaneous frequency conversion and beam steering, the space-time-modulated structure opens pathways for implementing multifunctional metasurfaces, enabling dynamic control over wave propagation, spectral manipulation, and spatial multiplexing. These results underline the potential of space-time-modulated Josephson junction arrays for next-generation electromagnetic systems, including quantum technologies, wireless communication networks, and spectral imaging systems.

\section{Conclusion}\label{sec:conc}
This study advances our understanding of dynamic Josephson junctions, a largely unexplored domain, by investigating their interaction with electromagnetic waves under space-time modulation. Through a rigorous mathematical framework, we modeled the propagation of electric and magnetic fields within and beyond arrays of space-time-modulated Josephson junctions, unveiling unique behaviors facilitated by dynamic modulation. Our findings demonstrate the potential of these arrays to enable sophisticated four-dimensional light manipulation, particularly through angular-frequency beam multiplexing that combines frequency conversion and beam-splitting functionalities. These capabilities pave the way for innovative applications in electromagnetic field engineering, ranging from superconducting quantum technologies to next-generation wireless communications, biomedical sensing, and radar systems. By bridging the knowledge gap in dynamic Josephson junctions and showcasing their transformative potential, this work lays a foundation for future research into dynamic superconducting systems and their applications in cutting-edge technologies.

\section{Appendix A\\Condition for the Floquet-Bloch Expansion}\label{sec:Append}
The wave equation inside the array reads
\begin{subequations}
\begin{equation}
	\nabla^2 \mathbf{H}_\text{s}(x,z,t) - \frac{1}{{{c^2}}}\frac{{{\partial ^2} \left[\mu_\text{s} (z,t)\mathbf{H}_\text{s}(x,z,t) \right]}}{{\partial {t^2}}}=0.
	\label{eqa:wave_eq}
\end{equation}
and then expand the wave equation within the modulated slab, as
\begin{equation}\label{eqa:A-wave_eq}
	\begin{split}
	c^2 &\frac{{{\partial ^2}\mathbf{H}_\text{s}}}{{\partial {x^2}}}+c^2 \frac{{{\partial ^2}\mathbf{H}_\text{s}}}{{\partial {z^2}}} = \frac{{{\partial ^2} \left[\mu_\text{s} (z,t)\mathbf{H}_\text{s} \right]}}{{\partial {t^2}}}\\
	&= \mathbf{H}_\text{s} \frac{{{\partial ^2} \mu_\text{s}(z,t) }}{{\partial {t^2}}} +\mu_\text{s}(z,t) \frac{{{\partial ^2} \mathbf{H}_\text{s} }}{{\partial {t^2}}} +2\frac{{{\partial } \mu_\text{s}(z,t) }}{{\partial {t}}} \frac{{{\partial \mathbf{H}_\text{s} } }}{{\partial {t}}},
	\end{split}
\end{equation}
\end{subequations}
\noindent where $\mathbf{H}_\text{s}=\mathbf{H}_\text{s}(x,z,t)$. Next, we apply the moving medium coordinate transformation
\begin{equation}\label{eqa:A-cord_transf}
	x'=x, \qquad    u=-\kappa_\text{s}z+\omega_\text{s}t, \qquad  t'=t.
\end{equation}
Next we express the partial derivatives in~\eqref{eqa:A-wave_eq} in terms of the new variable in~\eqref{eqa:A-cord_transf}, i.e.
\begin{subequations}\label{eqa:A-wave_transf}
	\begin{equation}
		\frac{\partial }{\partial x}=\frac{\partial {x'}}{\partial x} \frac{\partial }{\partial x'}+\frac{\partial u}{\partial x} \frac{\partial }{\partial u}+\frac{\partial t'}{\partial x} \frac{\partial }{\partial t'}=  \frac{\partial }{\partial x'},
	\end{equation}
	\begin{equation}
		\frac{\partial }{\partial z}=\frac{\partial {x'}}{\partial z} \frac{\partial }{\partial x'}+\frac{\partial {u}}{\partial z} \frac{\partial }{\partial u}+\frac{\partial t'}{\partial z} \frac{\partial }{\partial t'}= -\kappa_\text{s} \frac{\partial }{\partial u},
	\end{equation}
	\begin{equation}
		\frac{\partial^{2} }{\partial z^{2}}= \frac{\partial }{\partial z}\left(\frac{\partial}{\partial z}\right) =  \kappa_\text{s}^2 \frac{\partial^2 }{\partial u^2},
	\end{equation}
	\begin{equation}
		\frac{\partial }{\partial t}= \frac{\partial x'}{\partial t} \frac{\partial }{\partial x'}  +\frac{\partial u}{\partial t} \frac{\partial }{\partial u}+\frac{\partial t'}{\partial t} \frac{\partial }{\partial t'} = \frac{\partial }{\partial t'}+\omega_\text{s} \frac{\partial }{\partial u},
	\end{equation}
	\begin{equation}
			\begin{split}
		\frac{\partial^{2} }{\partial t^{2}}&= \left(\frac{\partial {t'}}{\partial t}\right)^2 \frac{\partial^2 }{\partial t'^2}+\left(\frac{\partial u}{\partial t}\right)^2 \frac{\partial^2 }{\partial u^2}+2 \left(\frac{\partial t'}{\partial t}\right) \frac{\partial u}{\partial t} \frac{\partial^2 }{\partial u \partial t'}\\& = \frac{\partial^2 }{\partial t'^2}+\omega_\text{s}^2 \frac{\partial^2 }{\partial u^2}+2 \omega_\text{s} \frac{\partial^2 }{\partial u \partial t'}.
			\end{split}
	\end{equation}
\end{subequations}
Using~\eqref{eqa:A-wave_transf} wherever appropriate, the different terms of~\eqref{eqa:A-wave_eq} become
\begin{subequations}\label{eqa:A-new_deriv}
	\begin{equation}
		c^2 \frac{{{\partial ^2} \mathbf{H}_\text{s}(x,z,t) }}{{\partial {x^2}}}= c^2 \frac{{{\partial ^2} \mathbf{H}_\text{s}(x',u,t') }}{{\partial {x'^2}}},
	\end{equation}
	\begin{equation}
		c^2 \frac{{{\partial ^2} \mathbf{H}_\text{s}(x,z,t) }}{{\partial {z^2}}}= c^2 \kappa_\text{s}^2 \frac{{{\partial ^2} \mathbf{H}_\text{s}(x',u,t') }}{{\partial {u^2}}},
	\end{equation}
	\begin{equation}
			\begin{split}
		\mathbf{H}_\text{s}(x,z,t) \frac{{{\partial ^2} \mu_\text{s}(z,t) }}{{\partial {t^2}}}&=\mathbf{H}_\text{s}(x',u,t')  \frac{\partial^2 }{\partial t'^2} \left(\sum\limits_{m = - \infty }^\infty  \tilde{\mu}_m e^{j m u}\right)\\&=-\omega_\text{s}^2 \mathbf{H}_\text{s}(x',u,t')  \sum\limits_{\substack{m =  - \infty\\m \neq  0}} ^\infty  m^2 \tilde{\mu}_m e^{j m u},
			\end{split}
	\end{equation}
	\begin{equation}
			\begin{split}
		\mu_\text{s}(z,t) \frac{{{\partial ^2} \mathbf{H}_\text{s}(x,z,t) }}{{\partial {t^2}}}=\sum\limits_{m = - \infty }^\infty  \tilde{\mu}_m e^{j m u} \Big( \frac{\partial^2 \mathbf{H}_\text{s}(x',u,t')}{\partial t'^2}\\+\omega_\text{s}^2 \frac{\partial^2 \mathbf{H}_\text{s}(x',u,t') }{\partial u^2}+2 \omega_\text{s} \frac{\partial^2 \mathbf{H}_\text{s}(x',u,t')}{\partial u \partial t'}  \Big),
			\end{split}
	\end{equation}
	\begin{equation}
		\begin{split}
		&	2\frac{{{\partial } \mu_\text{s}(z,t) }}{{\partial {t}}} \frac{{{\partial \mathbf{H}_\text{s}(x,z,t) } }}{{\partial {t}}}\\&= 2 \omega_\text{s} \frac{\partial }{\partial u} \sum\limits_{m = - \infty }^\infty  \tilde{\mu}_m e^{j m u}    \left( \frac{\partial \mathbf{H}_\text{s}(x',u,t') }{\partial t'}+\omega_\text{s} \frac{\partial \mathbf{H}_\text{s}(x',u,t') }{\partial u}  \right) \\&=2 j  \omega_\text{s} \sum\limits_{\substack{m =  - \infty\\m \neq  0} }^\infty  m \tilde{\mu}_m e^{j m u}    \left( \frac{\partial \mathbf{H}_\text{s}(x',u,t') }{\partial t'}+\omega_\text{s} \frac{\partial \mathbf{H}_\text{s}(x',u,t') }{\partial u}  \right),
		\end{split}
	\end{equation}

Grouping~\eqref{eqa:A-new_deriv} according to~\eqref{eqa:A-wave_eq} yields then the wave equation in terms of $x',u,t'$ and $\tilde{\mu}_m$:
\begin{equation}
	\begin{split}
	\Big(c^2 	&\kappa_\text{s}^2 - \omega_\text{s}^2 \sum\limits_{m = - \infty }^\infty  \tilde{\mu}_m e^{j m u}  \Big)  \frac{{{\partial ^2} \mathbf{H}_\text{s}(x',u,t') }}{{\partial {u^2}}} +c^2 \frac{{{\partial ^2} \mathbf{H}_\text{s}(x',u,t') }}{{\partial {x'^2}}} \\&-  \sum\limits_{m = - \infty }^\infty  \tilde{\mu}_m e^{j m u} \frac{\partial^2 \mathbf{H}_\text{s}(x',u,t')}{\partial t'^2}  -2 \omega_\text{s} \sum\limits_{m = - \infty }^\infty  \tilde{\mu}_m e^{j m u} \\&\cdot \frac{\partial^2 \mathbf{H}_\text{s}(x',u,t')}{\partial u \partial t'}  -2 j \omega_\text{s} \sum\limits_{\substack{m =  - \infty\\m \neq  0}}^\infty  m  \tilde{\mu}_m e^{j m u}    \Big( \frac{\partial \mathbf{H}_\text{s}(x',u,t') }{\partial t'} \\&+\omega_\text{s} \frac{\partial \mathbf{H}_\text{s}(x,z,t) }{\partial u}  \Big) + \omega_\text{s}^2 \sum\limits_{\substack{m =  - \infty\\m \neq  0}} ^\infty  m^2 \tilde{\mu}_m e^{j m u} \mathbf{H}_\text{s}(x',u,t')=0.
	\end{split}
	\label{eqa:A-long_eq_u_tp}
\end{equation}
\end{subequations}
For this equation to really represent the wave equation, it must maintain all of its order derivatives. This is generally the case, except when the coefficient of the first term vanishes, i.e.
\begin{subequations}
\begin{equation}
	c^2 \kappa_\text{s}^2 - \omega_\text{s}^2 \tilde{\mu}_0 - \omega_\text{s}^2 \sum\limits_{\substack{m =  - \infty\\m \neq  0}}^\infty  \tilde{\mu}_m e^{j m u}=0,
	\label{eqa:A-ggge}
\end{equation}
or, assuming a real permeability and hence $\Im\left\{\sum\limits_{m = - \infty }^\infty  \tilde{\mu}_m e^{j m u}\right\}=0$,
\begin{equation}
	\sum\limits_{\substack{m =  - \infty\\m \neq  0}}^\infty  g_m e^{-j m u}= \frac{\gamma^2  }{\mu_\text{r}}   -  g_0,
	\label{eqa:A-summ_sonic}
\end{equation}
\end{subequations}
\noindent where (2), (3) and (4) have been used. Assuming that the permeability variation is bounded between unity and infinity, i.e., \mbox{$ \bigg|\sum\limits_{\substack{m =  - \infty\\m \neq  0}}^\infty  g_m e^{-j m u}\bigg| \leq G_{\mu}  $}. Considering that $u$ is real, the condition~\eqref{eqa:A-summ_sonic} reduces to
\begin{subequations}
\begin{equation}
   \left| \frac{\gamma^2  }{\mu_\text{r}}   -  g_0\right|\leq G_{\mu},
	\label{eqa:A-son_cond}
\end{equation}

This indicates that the solution for $\mathbf{H}_\text{s}$ using the Floquet-Bloch expansion is not valid for the region

\begin{equation}\label{eqa:A-sonic_sin}
	\sqrt{\mu_\text{r} (g_0-G_{\mu} )} \leq 	\gamma  \leq
	\sqrt{\mu_\text{r} (g_0+G_{\mu} )}
\end{equation}
\end{subequations}

This interval is analogous to the sonic regime interval in aerodynamics, where the speed of an airplane matches or exceeds the velocity of sound, corresponding to the transonic and supersonic regimes, respectively. In this sonic interval, the standard space-time Floquet-Bloch decomposition fails to converge due to the unique interplay between wave propagation and the dynamic modulation of the medium. This breakdown arises because the energy and momentum conservation laws inherent to Floquet-Bloch theory are disrupted by the strong coupling between harmonics induced by the near-sonic modulation velocities.

\bibliographystyle{IEEEtran}
\bibliography{Taravati_Reference.bib}

\end{document}